\documentclass[preprints,article,accept,moreauthors,pdftex]{Definitions/mdpi} 

\firstpage{1} 
\pubvolume{1}
\issuenum{1}
\articlenumber{0}
\pubyear{2021}
\copyrightyear{2020}
\externaleditor{Academic Editors: Jos\'e A. de Freitas Pacheco and Pasquale Bosso} 
\datereceived{17 August 2021} 
\dateaccepted{10 September 2021} 
\datepublished{12 September 2021} 
\hreflink{https://doi.org/10.3390/sym13091678} 

\Title{Mass and rate of hierarchical black hole mergers in young, globular and nuclear star clusters}

\TitleCitation{Rates of hierarchical BH mergers}

\Author{Michela Mapelli $^{1,2,3}$\orcidA{}, Filippo Santoliquido $^{1,2}$\orcidB{}, Yann Bouffanais$^{1,2}$\orcidC{}, Manuel Arca Sedda$^4$\orcidD{}, M. Celeste Artale$^5$\orcidF{} and Alessandro Ballone$^{1,2,3}$\orcidG{}*}

\AuthorNames{Michela Mapelli, Filippo Santoliquido, Yann Bouffanais, Manuel Arca Sedda, M. Celeste Artale and Alessandro Ballone}

\AuthorCitation{Mapelli, M.; Santoliquido, F.; Bouffanais, Y.; Arca Sedda, M.; Artale, M. C.; Ballone, A.}

\address{%
$^{1}$  Physics and Astronomy Department Galileo Galilei, University of Padova, Vicolo dell'Osservatorio 3, I--35122, Padova, Italy\\
  $^2$  INFN-Padova, Via Marzolo 8, I--35131 Padova, Italy\\
  $^3$  INAF--Osservatorio Astronomico di Padova, Vicolo dell'Osservatorio 5, I--35122, Padova, Italy\\
  $^4$  Zentrum f\"{u}r Astronomie der Universit\"{a}t  Heidelberg, Astronomisches Rechen-Institut,M\"onchhofstrasse 12-14,Heidelberg, D-69120, DE\\
  $^5$  Institut f{\"u}r  Astro- und Teilchenphysik, Universit{\"a}t Innsbruck, Technikerstrasse 25/8, A-6020, Innsbruck, {\"O}sterreich
}

\corres{Correspondence: michela.mapelli@unipd.it}

\abstract{  Hierarchical mergers are one of the distinctive signatures of binary black hole (BBH) formation through dynamical evolution. Here, we present a fast semi-analytic approach to simulate hierarchical mergers in nuclear star clusters (NSCs), globular clusters (GCs) and young star clusters (YSCs). Hierarchical mergers are 
more common in NSCs than they are in both GCs and YSCs, because of the different escape velocity. 
The mass distribution of hierarchical BBHs strongly depends on the properties of first-generation BBHs, such as their progenitor's metallicity. 
  In our fiducial model, we form black holes (BHs) with masses up to $\sim{}10^3$ M$_\odot$ in NSCs and up to $\sim{}10^2$ M$_\odot$ in both GCs and YSCs. When escape velocities in excess of 100 km~s$^{-1}$ are considered, BHs with mass $>10^3$ M$_\odot$ are allowed to form in NSCs. Hierarchical mergers lead to the formation of BHs in the pair instability mass gap and intermediate-mass BHs, but only in metal-poor environments. 
  The local BBH merger rate in our models ranges from $\sim{}10$ to $\sim{} 60$ Gpc$^{-3}$ yr$^{-1}$; hierarchical BBHs in NSCs account for  $\sim{}10^{-2}- 0.2$ Gpc$^{-3}$ yr$^{-1}$, with a strong upper limit of $\sim{}10$ Gpc$^{-3}$ yr$^{-1}$. When comparing our models with the second gravitational-wave transient catalog, we find that multiple formation channels are favored to reproduce the observed BBH population.}


\begin{document}
\section{Introduction} \label{sec:intro}

The past six years have witnessed the first three observing runs of the Advanced LIGO and Virgo gravitational wave (GW) interferometers \citep{VIRGOdetector,LIGOdetector}, leading to the detection of about 50 binary compact object mergers \citep{abbottGW150914,abbottastrophysics,abbottO1,abbottGW170817,abbottO2,abbottO2popandrate,abbottO3a,abbottO3apopandrate,abbottGWTC-2.1}. This growing sample of GW observations represents a ``Rosetta stone'' to investigate the formation of binary compact objects.

Several channels can lead to the formation of binary black holes (BBHs): pairing of primordial black holes (e.g. \cite{carr1974,carr2016,bird2016}), binary star evolution through common envelope (e.g. \cite{tutukov1973, bethe1998,portegieszwart1998,belczynski2002, voss2003, podsiadlowski2004, belczynski2008,dominik2012, dominik2013,  mennekens2014, belczynski2016,eldridge2016,stevenson2017,mapelli2017,mapelli2018,giacobbo2018, giacobbo2018b, klencki2018,kruckow2018,spera2019,mapelli2019,neijssel2019,eldridge2019,tang2019}) or via homogeneous mixing (e.g. \cite{marchant2016,demink2016,mandel2016,dubuisson2020}), dynamical processes in triples (e.g. \cite{antonini2016b,antonini2017,arcasedda2018b,fragione2019,fragione2020}), young/open star clusters (YSCs, e.g. \cite{banerjee2010,ziosi2014,mapelli2016,askar2017,banerjee2017,banerjee2018,banerjee2020,dicarlo2019,dicarlo2020,kumamoto2019,kumamoto2020}), globular clusters (GCs, e.g. \cite{portegieszwart2000,downing2010,rodriguez2015,rodriguez2016,rodriguez2018,samsing2014,samsing2018,fragione2018,zevin2019,antonini2020}), nuclear star clusters (NSCs, e.g. \cite{oleary2009,miller2009,antonini2016,petrovich2017,rasskazov2019,arcasedda2018,arcasedda2019,arcasedda2020,arcasedda2020b}) and AGN disks (e.g. \cite{mckernan2012,mckernan2018,bartos2017,stone2017,yang2019,tagawa2020}).

One of the distinctive signatures of the dynamical scenario is the formation of hierarchical mergers, i.e. repeated mergers of stellar-origin black holes (BHs) that build up more massive ones \citep{miller2002,fishbach2017,gerosa2017,doctor2020,kimball2020}. This process is possible only in dense star clusters, where the merger remnant, which is initially a single BH, can acquire a companion by dynamical exchanges \citep{hills1980}.   The main obstacle to the formation of second-generation (2g) BHs via hierarchical mergers is the high relativistic kick that the merger remnant receives at birth, because of radiation
of linear momentum through beamed GW emission (e.g. \cite{fitchett1983,favata2004,campanelli2007,lousto2011}). This kick can be up to several thousand km s$^{-1}$ and can easily eject the BH remnant from its parent star cluster \citep{holley-bockelmann2008,moody2009,fragione2018b,gerosa2019,arcasedda2020}. Hence, the interplay between the properties of the host star cluster (e.g. its escape velocity), those of the first-generation (1g) BBH population and the magnitude of the kick decides the maximum mass of a merger remnant in a given environment. This might be used to constrain the formation channels of BBHs. 

The spins of 1g BHs are one of the critical ingredients, because relativistic kicks are sensitive  to spin magnitudes and orientation (e.g. \cite{lousto2012,maggiore2018}). In the zero-spin assumption, more than 10\% of merging BBHs from GCs have components formed from previous mergers, accounting for more than 20\% of the mergers from GCs detectable by Advanced LIGO and Virgo \citep{rodriguez2019}. 

Due to their high escape velocity ($v_{\rm esc}\sim{}100$~km~s$^{-1}$), NSCs are more likely to retain hierarchical mergers than other star clusters (e.g. \cite{antonini2016,yang2019,arcasedda2019c,arcasedda2020}). \cite{antonini2019} recently found that BH growth becomes substantial for $v_{\rm esc}>300$ km s$^{-1}$, leading to the formation of intermediate-mass BHs (IMBHs, see also \cite{fragione2020}).
Hence, hierarchical mergers can build up IMBHs and also partially fill the pair instability 
mass gap between $\sim{}60$ and $\sim{}120$ M$_\odot$ \citep{belczynski2016pair,woosley2017,spera2017,farmer2019,mapelli2020,renzo2020}. For this reason, hierarchical mergers are one of the most likely formation scenarios for GW190521 \citep{abbottGW190521,abbottGW190521astro}, as already explored by several authors (\cite{kimball2020a,fragione2020b,rizzuto2020,liulai2021,mapelli2021,dallamico2021}, but see \cite{dicarlo2019,dicarlo2020,kremer2020,roupas2019,rice2020,safarzadeh2020,palmese2020,belczynski190521,farrell2020,tanikawa2020,costa2021,deluca2021} for other possible scenarios).











The main challenge of studying hierarchical mergers is the computational cost. It is nearly impossible to investigate the relevant parameter space with hybrid Monte Carlo and/or N-body simulations of star clusters, especially GCs and NSCs. Here, we present a new fast and flexible semi-analytic model to investigate hierarchical mergers in different environments, complementary to dynamical simulations. Our new tool allows us to probe the parameter space (1g masses, spins, delay times, 2g masses, spins and delay times, escape velocity from the parent cluster and kick magnitudes) and to reconstruct the merger rate evolution of each formation channel, with just minimal model assumptions. 

\section{Methods} \label{sec:methods}

We consider four different environments: i) the field, where hierarchical mergers are not possible, ii) young star clusters (YSCs), which are the main birth site of massive stars in the local Universe (e.g. \cite{portegieszwart2010}), iii) globular clusters (GCs), and iv) nuclear star clusters (NSCs). To evaluate the properties of 1g mergers, we start from catalogs of single and binary BHs obtained with population-synthesis simulations. When the 1g BHs merge, we estimate the relativistic kick $v_{\rm kick}$ and the escape velocity from the parent star cluster $v_{\rm esc}$. If $v_{\rm kick}<v_{\rm esc}$, we assume that the merger remnant remains bound to its parent star cluster and can pair with another BH dynamically. We estimate the mass and spin of the merger remnant and of its new companion, as detailed below.  Then, we randomly draw a new delay time between previous and next merger. If the sum of the new delay time and the previous one is  shorter than the Hubble time, we repeat the loop for another generation.

\subsection{First generation (1g) mergers}\label{sec:firstgen}

We take the mass of 1g BHs from our population synthesis simulations. In particular, we used our code {\sc mobse} \citep{mapelli2017,giacobbo2018,giacobbo2018b}. {\sc mobse} 
is an upgraded and customized version of {\sc bse} \citep{hurley2002}.  The treatment of stellar winds is one of the key aspects affecting the final mass of BHs and is subject to a number of uncertainties \cite{smith2014}. In {\sc mobse}, mass loss by stellar winds for massive host stars (O-type, B-type, luminous blue variable and Wolf-Rayet stars) is modeled as $\dot{M}\propto{}Z^\beta$, where  $Z$ is the metallicity and
  \begin{equation}\label{eq:gamma}
    \beta=\left\{
      \begin{array}{ll}
      0.85, & \text{if}\ \Gamma_{e} \leq{} 2/3 \\
      2.45-2.4\Gamma_{e}, & \text{if}\  2/3<\Gamma_{e} \leq{} 1\\
      0.05, & \text{if}\ \Gamma_{e} > 1
      \end{array}
    \right.
  \end{equation}
In eq.~\ref{eq:gamma},   $\Gamma_e$ is the Eddington ratio, i.e. the ratio between the luminosity of the star and its Eddington value. This formalism, introduced by \cite{chen2015}, is a fit to the models presented in \cite{graefener2008}. It accounts  for both  the $Z$ dependence of line-driven winds \citep{vink2001} and  the importance of Thomson scattering when a star is nearly radiation pressure dominated \citep{graefener2008, vink2011}.

In detail, for O and B-type stars with  effective temperature $T_{\rm eff}\ge{}12500$ K, we use the same fitting formulas as introduced by \cite{vink2001}, but we correct the $Z$ dependence as described in equation~\ref{eq:gamma}. We model the mass loss rate of Wolf-Rayet stars as \citep{belczynski2010}
\begin{equation}
\dot{M}=10^{-13}\,{}L^{1.5}\,{}\left(\frac{Z}{{\rm Z}_\odot}\right)^{\beta}\,{}{\rm M}_\odot\,{}{\rm yr}^{-1}.
    \end{equation}
The mass loss rate of luminous blue variable stars is \citep{giacobbo2018}
\begin{equation}\label{eq:lbv}
\dot{M}=1.5\times{}10^{-4}\,{}\left(\frac{Z}{{\rm Z}_\odot}\right)^{\beta}\,{}{\rm M}_\odot\,{}{\rm yr}^{-1}.
    \end{equation}
    Finally, the treatment of mass loss of cold massive stars is the same as originally described by \cite{hurley2000}.

The effect of core-collapse supernovae on the mass of compact objects is described following the delayed model of \cite{fryer2012}. 
According to this model, stars with final carbon-oxygen mass $m_{\rm CO}\gtrsim{}11$ M$_\odot$ collapse to a BH directly. The minimum BH mass is 3 M$_\odot$. Following \cite{timmes1996} and \cite{zevin2020}, we compute neutrino mass loss for both neutron stars and BHs  as
\begin{equation}
    m_\nu{}=\min\left[\frac{\left(\sqrt{1+0.3\,{}m_{\rm bar}}-1\right)}{0.15},\,{}0.5\,{}{\rm M}_\odot\right],
\end{equation}  
where  $m_{\rm bar}$ is the baryonic mass of the compact object.       The resulting gravitational mass of the compact object is $m_{\rm grav} = m_{\rm bar} - m_\nu{}$.

      Stars with helium core mass (at the end of carbon burning) $32\le{}m_{\rm He}\le{}64$ and $64\le{}m_{\rm He}\le{}135$ undergo pulsational pair instability and pair instability supernovae, respectively \citep{woosley2017}. Stars that undergo a pair instability supernova leave no compact remnant, while stars going through pulsational pair instability become BHs with mass $m_{\rm BH}=\alpha{}_{\rm P}\,{}m_{\rm no,\,{}PPI}$, where the possible values of $\alpha{}_{\rm P}\leq{}1$ are discussed in \cite{mapelli2020} and $m_{\rm no,\,{}PPI}$ is the BH mass from direct collapse, if pulsational pair instability is not accounted for. 
      Finally, electron-capture supernovae are included following \cite{giacobbo2018c}. For natal kicks, we adopt the prescription $v_{\rm k}\propto{}m_{\rm ej}\,{}m_{\rm rem}^{-1}$, where $m_{\rm ej}$ is the mass of the ejecta and $m_{\rm rem}$ is the mass of the compact remnant (neutron star or BH,  \cite{giacobbo2020}).

Binary evolution processes (wind mass transfer, Roche lobe overflow, common envelope, mergers, tidal evolution, GW decays) are implemented as in \cite{hurley2002}, with one significant exception. During Roche lobe overflow, the accretion rate  is calculated as
  \begin{equation}\label{eq:MT}
    \dot{m}_a=\left\{
    \begin{array}{ll}
      f_{\rm MT}\,{}|\dot{m}_d| & \textrm{if non-degenerate accretor}\\
      \min{(f_{\rm MT}\,{}|\dot{m}_d|,\dot{m}_{\rm Edd})} & \text{otherwise},
    \end{array}
    \right.
  \end{equation}
  where $\dot{m}_a$ is the accretion rate, $\dot{m}_d$ is the mass loss rate by the donor, $\dot{m}_{\rm Edd}$ is the Eddington accretion rate and $f_{\rm MT}\in{}(0,\,{}1]$ is the accretion efficiency. Here, we consider $f_{\rm MT}=0.1,\,{}0.5,\,{}1.0$. The original prescriptions by \cite{hurley2002} are close to $f_{\rm MT}=1.0$.    We parametrize common envelope evolution with the parameter $\alpha{}$ \citep{hurley2002}. Here, we consider $\alpha{}=1,\,{}5,\,{}10$, large values of $\alpha{}$ meaning that the envelope is easily ejected, without much shrinking of the binary. In its original meaning \citep{webbink1984}, $\alpha{}$ is the fraction of orbital energy that is transferred to the envelope during the spiral-in phase. Here, we also consider values of $\alpha{}>1$, because the original formalism     does not include additional contributions to the energy budget (e.g. \cite{ivanova2013,fragos2019}). 

Giacobbo et al. (2018, \cite{giacobbo2018}) have shown (e.g. their Figure~4) that with these prescriptions for stellar and binary evolution the maximum mass of a single BH can be as high as $m_{\rm BH}\approx{}65-70$ M$_\odot$. Such massive BHs come from metal-poor stars\footnote{A metallicity $Z\sim{0.0002}\approx{0.01}{\rm Z}_\odot$ is typical of population~II stars. The most metal poor GCs in the Milky Way have metallicity $Z\sim{0.0002}$  \cite{gratton2004}. One of the most metal-poor dwarf galaxies in the local Universe, IZw18, has metallicity $Z\approx{0.0004}$ \cite{fiorentino2010}.} ($Z\sim{}0.0002$) with initial mass $m_{\rm ZAMS}\approx{}70-80$ M$_\odot$, which retain most of their hydrogen envelope at the time of collapse and have sufficiently small helium cores to avoid pulsational pair instability  \citep{mapelli2020}. However, the maximum mass of a BH merging within a Hubble time as a result of isolated binary evolution is only $m_{\rm BH}\approx{}50$ M$_\odot$ \citep{giacobbo2018}. This happens because binary stars that are sufficiently tight to merge within a Hubble time by GW emission evolve through mass transfer and common envelope. These processes remove the hydrogen envelope, leading to smaller BH masses. Hence, the resulting BBH cannot have a total mass higher than $m_{\rm TOT}\approx{}100$ M$_\odot$.

In dynamical environments, exchanges and dynamical hardening might allow even more massive BHs to merge, up to total binary masses $m_{\rm TOT}\approx{}130-140$ M$_\odot$ \citep{dicarlo2020}. For this reason, we consider two different sets of models for 1g masses. In our fiducial model A5F05 (conservative approach), the masses of 1g BBHs are randomly drawn from catalogs of BBHs simulated with {\sc mobse}. In this case, the evolution of the semi-major axis $\mathcal{A}$ and of the eccentricity $e$ of the BBHs are calculated as \citep{peters1964}
\begin{eqnarray}\label{eq:peters1964}
  \frac{{\rm d}\mathcal{A}}{{\rm d}t}=-\frac{64}{5}\,{} \frac{G^3 \,{} m_1 \,{} m_2 \,{} (m_1+m_2)}{c^5 \,{} \mathcal{A}^3\,{} (1-e^2)^{7/2}} \,{} \left(1+\frac{73}{24}\,{}e^2+\frac{37}{96}\,{} e^4\right)\nonumber\\
  \frac{{\rm d}e}{{\rm d}t}=-\frac{304}{15}\,{} e \frac{ G^3 \,{} m_1 \,{} m_2 \,{} (m_1+m_2)}{c^5 \,{}\mathcal{A}^4 \,{}  (1-e^2)^{5/2}}\,{}\left(1+\frac{121}{304} \,{} e^2\right), 
\end{eqnarray}
where $G$ is the gravity constant, $c$ is the speed of light, $m_1$ and $m_2$ are the masses of the primary and secondary BH, respectively.

In the HIGH\_MASS model (optimistic approach), the masses of field BBHs are still taken from catalogs of BBH mergers, while the masses of 1g dynamical BBHs 
are uniformly drawn from the list of all the BHs formed with {\sc mobse}, which include both single and binary BHs, both merging and non-merging systems. This ensures that the masses of dynamically formed 1g BBHs can reach $m_{\rm TOT}\approx{}140$ M$_\odot$, while the maximum total mass of field binaries is $m_{\rm TOT}\approx{}100$ M$_\odot$. In the HIGH\_MASS case, we randomly pair the primary and the secondary component and we randomly draw the delay time (i.e., the time elapsed from the formation of the BBH to its merger) from a distribution $dN/dt\propto{}t^{-1}$ between $t_{\rm min}=10^7$ yr and $t_{\rm max}=1.4\times{}10^{10}$ yr \citep{dominik2012,dicarlo2020}.

We define the dimensionless spin magnitude $a$ as $a\equiv{}\mathcal{S}\,{}c/(G\,{}m_{\rm BH}^2)$, where $\mathcal{S}$ is the spin magnitude in physical units. Spin magnitudes of 1g BHs are randomly drawn from a Maxwellian distribution with fiducial one-dimension root-mean square $\sigma_a=0.2$ and truncated at $a=1$. We consider also two extreme cases in which $\sigma_a=0.01$ (LOW\_SPIN model) and $\sigma_a=0.4$ (HIGH\_SPIN model). This is just a toy model because the uncertainties on BH spin magnitudes from stellar evolution and core-collapse supernova models are still too large to make predictive statements. Angular momentum transport via the magnetic Tayler-Spruit instability might be effective and lead to predominantly low spins (e.g. \cite{fuller2019,belczynski2020}), while binary evolution processes can significantly affect the overall picture \citep{qin2018,qin2019}. Our LOW\_SPIN case can be interpreted as the result of the  spin distribution inferred by \cite{fuller2019}. Spin directions in dynamical BBHs are isotropically distributed over a sphere \citep{rodriguez2016}.

Our set of runs is described in Table~\ref{tab:table1}. The initial {\sc mobse} population of each model is obtained running $1.2\times{}10^8$  binary stars with metallicity $Z=0.02$, 0.016, 0.012, 0.008, 0.006, 0.004, 0.002, 0.0016, 0.0012, 0.0008, 0.0004, 0.0002. The initial mass of the primary is drawn from a Kroupa initial mass function \citep{kroupa2001} between 5 and 150 M$_\odot$. Mass ratios, orbital periods and eccentricities are randomly drawn following the distributions presented in \cite{sana2012}.

\begin{specialtable}[H] 
\caption{Main properties of the runs presented in this paper\label{tab:table1}}
\footnotesize{}
\begin{tabular}{cccccccc}
\toprule
\textbf{Run Name} & 
 \textbf{$\alpha{}$} & \textbf{$f_{\rm MT}$}  & \textbf{$t_{\rm del}$} & \textbf{$\sigma_a$}  & \textbf{$m_2$} & \textbf{$t_{\rm min}$ [Myr]} & \textbf{$\log_{10}{\left(v_{\rm esc}/{\rm km}\,{}{\rm s}^{-1}\right)}$} \\ 
\textbf{} & 
 \textbf{1g} & \textbf{1g} & \textbf{1g} & \textbf{1g} & \textbf{$N$g} & \textbf{$N$g} & \textbf{NSC,  GC,  YSC} \\
\midrule
Fiducial, A5F05 & 5.0 & 0.5 & eq.~\ref{eq:peters1964} & 0.2 & uniform  & 10 & $2\pm{}0.2$,  $1.3\pm{}0.2$,  $0.7\pm{}0.2$ \\ 
A5F01 & 5.0 & 0.1 & eq.~\ref{eq:peters1964} & 0.2 & uniform  & 10 & $2\pm{}0.2$,  $1.3\pm{}0.2$,  $0.7\pm{}0.2$ \\ 
A5F1 & 5.0 & 1.0 & eq.~\ref{eq:peters1964} & 0.2 & uniform  & 10 & $2\pm{}0.2$,  $1.3\pm{}0.2$,  $0.7\pm{}0.2$ \\ 
A1F01 & 1.0 & 0.1 & eq.~\ref{eq:peters1964} & 0.2 & uniform  & 10 & $2\pm{}0.2$,  $1.3\pm{}0.2$,  $0.7\pm{}0.2$ \\ 
A1F05 & 1.0 & 0.5 & eq.~\ref{eq:peters1964} & 0.2 & uniform  & 10 & $2\pm{}0.2$,  $1.3\pm{}0.2$,  $0.7\pm{}0.2$ \\ 
A1F1 & 1.0 & 1.0 & eq.~\ref{eq:peters1964} & 0.2 & uniform  & 10 \vspace{0.1cm} & $2\pm{}0.2$,  $1.3\pm{}0.2$,  $0.7\pm{}0.2$ \\ 
A10F01 & 10.0 & 0.1 & eq.~\ref{eq:peters1964} & 0.2 & uniform  & 10 & $2\pm{}0.2$,  $1.3\pm{}0.2$,  $0.7\pm{}0.2$ \\ 
A10F05 & 10.0 & 0.5 & eq.~\ref{eq:peters1964} & 0.2 & uniform  & 10 & $2\pm{}0.2$,  $1.3\pm{}0.2$,  $0.7\pm{}0.2$ \\ 
A10F1 & 10.0 & 1.0 & eq.~\ref{eq:peters1964} & 0.2 & uniform  & 10 & $2\pm{}0.2$,  $1.3\pm{}0.2$,  $0.7\pm{}0.2$ \vspace{0.1cm}\\ 
HIGH\_MASS & -- & -- & $t^{-1}$ & 0.2 & uniform   & 10 & $2\pm{}0.2$,  $1.3\pm{}0.2$,  $0.7\pm{}0.2$\vspace{0.1cm} \\ 
SMALL\_M2 & 5.0 & 0.5 & eq.~\ref{eq:peters1964} & 0.2 & {\sc mobse}  & 10 & $2\pm{}0.2$,  $1.3\pm{}0.2$,  $0.7\pm{}0.2$\vspace{0.1cm} \\ 
LOW\_SPIN & 5.0 & 0.5 & eq.~\ref{eq:peters1964} & 0.01 & uniform  & 10 & $2\pm{}0.2$,  $1.3\pm{}0.2$,  $0.7\pm{}0.2$ \\ 
HIGH\_SPIN & 5.0 & 0.5 & eq.~\ref{eq:peters1964} & 0.4 & uniform  & 10 & $2\pm{}0.2$,  $1.3\pm{}0.2$,  $0.7\pm{}0.2$\vspace{0.1cm} \\ 
SHORT\_DELAY & 5.0 & 0.5 & eq.~\ref{eq:peters1964} & 0.2 & uniform  & 0.1 & $2\pm{}0.2$,  $1.3\pm{}0.2$,  $0.7\pm{}0.2$ \\ 
LONG\_DELAY & 5.0 & 0.5 & eq.~\ref{eq:peters1964} & 0.2 & uniform  & 100 & $2\pm{}0.2$,  $1.3\pm{}0.2$,  $0.7\pm{}0.2$ \vspace{0.1cm}\\ 
BROAD\_VESC & 5.0 & 0.5 & eq.~\ref{eq:peters1964} & 0.2 & uniform  & 10 & $2\pm{}0.3$,  $1.3\pm{}0.3$,  $0.7\pm{}0.3$ \\ 
NARROW\_VESC & 5.0 & 0.5 & eq.~\ref{eq:peters1964} & 0.2 & uniform  & 10 & $2\pm{}0.1$,  $1.3\pm{}0.1$,  $0.7\pm{}0.1$ \vspace{0.1cm}\\ 
\bottomrule
\end{tabular}
\flushleft{Column 1: Name of the model.
  Column 2: parameter $\alpha{}$ of  common envelope for 1g BBHs. Column 3: parameter $f_{\rm MT}$ of accretion efficiency for non-degenerate accretors (eq.~\ref{eq:MT}) in the case of  1g BBHs. Column 4: delay time distribution of 1g BBHs; `eq.~\ref{eq:peters1964}' indicates that the delay times were calculated solving eq.~\ref{eq:peters1964}; `$t^{-1}$' means that delay times were randomly drawn from $dN/dt\propto{}t^{-1}$. Column 5: one-dimensional root-mean square associated with the Maxwellian distribution used to extract 1g spin magnitudes; we adopted values $\sigma_a=0.2$ (fiducial), 0.01 (LOW\_SPIN), 0.4 (HIGH\_SPIN). Column 6: distribution from which we drew the mass of the secondary component in the $N$g BBHs; `uniform' means that $m_2$ is uniformly distributed between $m_{\rm MIN}=3$ M$_\odot$ and $m_{\rm MAX}=m_1$ (fiducial), `{\sc mobse}' means that we randomly selected $m_2$ from catalogs of BHs simulated with {\sc mobse} (used in the SMALL\_M2 run). 
  Column~7: $t_{\rm min}$ is the minimum delay time for $N$g BBHs. Column~8: mean and standard deviation of the lognormal distribution of escape velocities $v_{\rm esc}$ for NSCs, GCs and YSCs.}
\end{specialtable}

\subsection{Relativistic kicks}

We model the magnitude of relativistic kicks according to equation~12 of \cite{lousto2012}:
\begin{equation}\label{eq:vk}
    v_{\rm kick}=\left(v_m^2+v_\perp^2+2\,{}v_m\,{}v_\perp\,{}\cos{\xi{}}+v_\parallel^2\right)^{1/2},    
\end{equation}
where
\begin{eqnarray}
  v_m=A\,{} \eta{}^2 \,{}\frac{(1-q)}{(1+q)}\,{} (1 + B\,{} \eta{})\nonumber\\
    v_\perp{}=H\,{}\frac{\eta^2}{(1+q)}\,{}\left|a_{1\parallel}-q\,{}a_{2\parallel}\right|\nonumber\\
    v_\parallel=\frac{16\,{}\eta^2}{(1+q)}\,{}\left[V_{1,1}+V_A\,{}S_\parallel+V_B\,{}S_\parallel^2+V_C\,{}S_\parallel^3\right]\nonumber{}\\
    \,{}\left|a_{1\perp{}}-q\,{}a_{2\perp{}}\right|\,{}\cos{\left(\phi_\Delta-\phi\right)}.    
\end{eqnarray}
In the above equations, $q=m_2/m_1$ with $m_2\le{}m_1$, $\eta=q\,{}(1+q)^{-2}$, $A=1.2\times{}10^4$~km~s$^{-1}$, $B=-0.93$, $H=6.9\times{}10^3$~km~s$^{-1}$, $(V_{1,1},\,{}V_A,\,{}V_B,\,{}V_C)=$~(3678, 2481, 1792, 1506) km s$^{-1}$, $\xi{}=145^\circ{}$ \citep{lousto2009}, while $\vec{a}_1$ and $\vec{a}_2$ are the spin vectors of the primary and secondary BHs, respectively. Moreover, $a_{1\parallel{}}$ ($a_{2\parallel}$) is the component of the spin of the primary (secondary) BH parallel to the orbital angular momentum of the binary system, while   $a_{1\perp{}}$ ($a_{2\perp}$) is the component of the spin of the primary (secondary) BH lying in the orbital plane. $S_\parallel$ is the component parallel to the orbital angular momentum of the vector $\vec{S}=2\,{}(\vec{a}_1+q^2\vec{a}_2)/(1+q)^2$. Finally, $\phi_\Delta$ represents the angle between the direction of the infall at merger (which we randomly draw in the BBH orbital plane) and the in-plane component of $\vec{\Delta} \equiv{} (m_1 + m_2 )^2\,{} \left(\vec{a}_1 - q \,{} \vec{a}_2 \right)/(1+q)$, while $\phi$  is the phase of the BBH, randomly drawn between 0 and $2\,{}\pi{}$. Equation~\ref{eq:vk} results from an empirical model for the recoil velocity as
a function of the progenitor’s parameters (mostly $q$, $\vec{a}_1$ and $\vec{a}_2$). The basic idea behind it is that the recoil of spinning
BHs is mostly produced close to the time of merger \cite{lousto2008}, and can be modeled by a
parametrized dependence of the leading (on spins and
mass ratio) post-Newtonian expressions for the linear momentum radiated \citep{kidder1995}. The term $v_{\rm m}$ mainly comes from the contribution of the mass ratio $q$ to the linear momentum radiated (the merger of an unequal mass BBH produces a kick even if the two BHs are non spinning), while $v_\perp$ and $v_\parallel$ account for the contribution of the spin components aligned and orthogonal to the orbital angular momentum, respectively. The coefficients in eq.~\ref{eq:vk} are given by fits to full numerical relativity simulations, as detailed in \cite{lousto2012}. 
This formalism yields kicks up to $\sim{4000}$ km s$^{-1}$, but the most common kicks are of the order of a few hundred km s$^{-1}$, as shown in Figure~9 of \cite{lousto2012} and discussed in Section~\ref{sec:properties}.

\begin{figure}[]
    \includegraphics[width = 10.5cm]{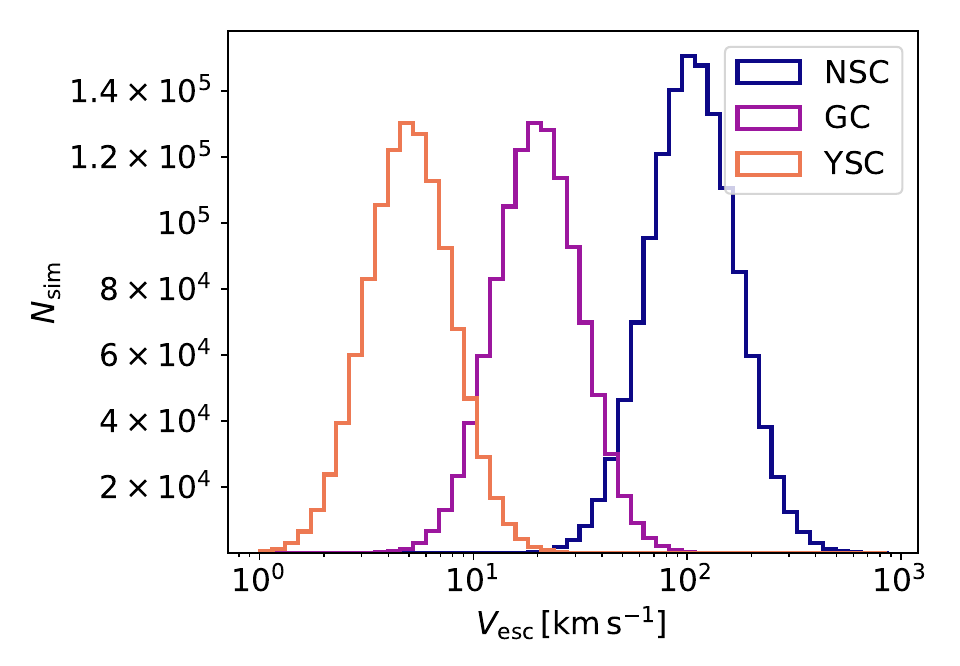}
\caption{Distribution of escape velocities adopted in the fiducial case. See Section~\ref{sec:vesc}.  \label{fig:vesc}}
\end{figure}

\subsection{Escape velocities}\label{sec:vesc}

For each merger, we calculate the relativistic kick magnitude as in eq.~\ref{eq:vk} and then  compare it with the escape velocity of the host cluster $v_{esc}$. If $v_{\rm kick}<v_{\rm esc}$, the remnant is retained inside its host cluster and can undergo another merger. Otherwise, it is ejected and remains a single BH. We randomly draw $v_{\rm esc}$ from a log-normal distribution with median $\langle{}\log_{10}(v_{esc}/{\rm km}\,{}{\rm s}^{-1})\rangle=2.0,$ 1.3, 0.7 (standard deviation $\sigma{}_{\rm v}=0.2,$ 0.2, 0.2) for NSCs, GCs, and YSCs, respectively. This choice is motivated by observations of  NSCs \citep{antonini2016}, GCs \citep{harris2013} and YSCs \citep{portegieszwart2010} in the local Universe. Figure~\ref{fig:vesc} shows the distribution of escape velocities in our fiducial case. 
In the next sections, we show what happens if we change these assumptions. Namely, in the BROAD\_VESC (NARROW\_VESC) model we assume $\sigma_{\rm v}=0.3$ (0.1) for NSCs, GCs and YSCs.

\subsection{\textit{N}th generation ($N$g) mass and spin}

We model the mass and spin of merger remnants using the fitting formulas in \cite{jimenez-forteza2017} for quasi-circular non-precessing mergers (see also \cite{rezzolla2008,hofmann2016,arcasedda2020}). 
 The final mass is $\approx{}0.95$ the total mass of the two merging BHs, while the final spin magnitude clusters around $a_{\rm f}\approx{}0.75$. If the merger remnant is retained, 
 it eventually pairs up with another BH. The mass of the companion is selected in two different ways.
To account for the fact that the secondary component might be either a 1g or a $N$g object with $N>1$, we uniformly draw the mass of the secondary BH, $m_2$, between $m_{\rm MIN}=3$ M$_\odot$ and $m_{\rm MAX}=m_1$ (fiducial model). This assumption favors $N$g$-N$g mergers with respect to $N$g$-1$g mergers. To account for cases in which the primary component is a $N$g merger (with $N>1$) and the secondary component is  a 1g BH, we draw the mass of the secondary BH from the population-synthesis catalogs of 1g BHs (model SMALL\_M2).

The spins of  secondary BHs are randomly drawn from a Maxwellian distribution with default one-dimensional root-mean square $\sigma_a=0.2$. In the LOW\_SPIN (HIGH\_SPIN) case, $\sigma{}_a=0.01$ (0.4). This is a simplification, because we do not distinguish whether the secondary component is a $N$g or a 1g BH. The spin vectors of both the primary and  secondary BH are isotropically distributed over a sphere. 

Finally, in all models we check that the mass of the remnant BH is always less than
\begin{equation}
m_{\rm th}=10^{-3}\,{}m_{\rm SC}\approx{}10^3\,{}{\rm M}_\odot\,{}\left(\frac{v_{\rm esc}}{100\,{}{\rm km}\,{}{\rm s}^{-1}}\right)^2.
\end{equation}
This condition is equivalent to assuming that the most massive BH cannot be more massive than the total mass of all BHs in the star cluster, assuming a Kroupa IMF. If a BH hits this mass threshold, it cannot grow any further by hierarchical merger.

\subsection{Delay times of $N$g mergers}

For $N$g mergers (where $N=2$ or more), 
we randomly draw the delay times according to a distribution uniform in $dN/dt\propto{}t^{-1}$ \citep{dominik2012} and spanning from $t_{\rm min}=[0.1,\,{}10,\,{}100]$ Myr (see column~7 of Table~\ref{tab:table1}) to $t_{\rm max}=1.4\times{}10^{4}$ Myr.  By adopting $dN/dt\propto{}t^{-1}$, we have assumed that GW decay is the dominant effect to determine the delay time. 
Since the GW timescale $\tau_{\rm GW}\propto{}\mathcal{A}^4$, where $\mathcal{A}$ is the initial semi-major axis of the BBH \citep{peters1964}, for an initial semi-major axis distribution $f(\mathcal{A})\propto{}\mathcal{A}^{-1}$, we obtain a delay time distribution $dN/dt\propto{}t^{-1}$ \citep{piran1992,totani2008,wang2020FRB}. 
This is the crudest assumption in our method, because we neglect the impact of dynamical hardening on the evolution of the BBH semi-major axis. However, this assumption is supported by $N-$body simulations of dense YSCs \citep{dicarlo2019,dicarlo2020}, which show that dynamical BBH mergers follow a trend $dN/dt\propto{}t^{-1}$. The choice of $t_{\rm min}$ depends on the time for dynamical pairing of the BBH $t_{\rm dyn}$, i.e. the time needed for a single BH to find a new companion BH via dynamical interactions. This can be estimated as the sum of the dynamical friction timescale ($t_{\rm DF}$, i.e. the time over which the merger remnant, which is ejected in the outskirts of the star cluster by the relativistic kick, sinks back to the core of the parent cluster by dynamical friction, \citep{chandrasekhar1943}) and the three-body timescale ($t_{\rm 3bb}$, i.e. the timescale for BBH formation by three-body encounters, \citep{lee1995}):
\begin{eqnarray}\label{eq:tDFt3bb}
t_{\rm DF}=9.5\,{}{\rm Myr}\,{}\left(\frac{m_{\rm BH}}{30\,{}M_\odot}\right)^{-1}\,{}\left(\frac{M_{\rm SC}}{10^6M_\odot}\right)\,{}\left(\frac{n}{10^6\,{}{\rm pc}^{-3}}\right)^{-1/2},\nonumber{}\\
t_{\rm 3bb}=0.1\,{}{\rm Myr}\,{}\left(\frac{n}{10^6\,{}{\rm pc}^{-3}}\right)^{-2}\,{}\left(\frac{\sigma{}_{\rm SC}}{30\,{}{\rm km}\,{}{\rm s}^{-1}}\right)^9\,{}\left(\frac{m_{\rm BH}}{30\,{}M_\odot}\right)^{-5},
\end{eqnarray}
where $M_{\rm SC}$ and $n$ are the star cluster mass and central number density, 
while $\sigma{}_{\rm SC}=v_{\rm esc}/(2\,{}\sqrt{3})$ is the one-dimensional velocity dispersion. In eq.~\ref{eq:tDFt3bb}, we  assumed that the average mass of a star in the star cluster is 1~M$_\odot$ and that, locally, the star cluster is in equipartition \citep{spitzer1987}. The timescale for  dynamical BBH formation  is then $t_{\rm dyn}=t_{\rm DF}+t_{\rm 3bb}$. For most star clusters, we find $0.1\leq{}t_{\rm dyn}/{\rm Myr}\leq{}100$, roughly corresponding to the values of $t_{\rm min}$ we assume in our analysis (column~7 of Table~\ref{tab:table1}).  

\subsection{Summary of the models}

In Table~\ref{tab:table1}, A5F05 is our fiducial model. Models with name A$i$F$j$ (with $i=1,\,{}5,\,{}10$ and $j=01,\,{}05,\,{}1$) differ from the fiducial model only for the choice of the common envelope parameter $\alpha{}$ ($\alpha=1,$ 5, 10 if $i=1,\,{}5,\,{}10$) and of the accretion efficiency $f_{\rm MT}$ ($f_{\rm MT}=0.1$, 0.5, 1 if $j=01,\,{}05,\,{}1$). The model HIGH\_MASS differs from the fiducial model for the choice of the masses of 1g BHs and for their delay time distribution. In the model HIGH\_MASS, 1g BH masses in star clusters are uniformly sampled from all BHs generated with {\sc mobse} (including single BHs), mimicking the impact of dynamical exchanges. Delay times are drawn from $dN/dt\propto{}t^{-1}$.

The SMALL\_M2 model differs from the fiducial one for the masses $m_2$ of secondary BHs in $N$g mergers, which are randomly drawn from 1g BHs. Hence, in this model all BBH mergers occur with a  1g secondary BH. In contrast, $m_2$ is uniformly sampled  in $[3\,{}{\rm M}_\odot,m_1]$ in all the other models.

The LOW\_SPIN and HIGH\_SPIN models differ from the fiducial model only for the distribution of spin magnitudes of 1g BHs. The one-dimensional root-mean square $\sigma_a$ is 0.01 and 0.4 in LOW\_SPIN and HIGH\_SPIN, respectively.

The SHORT\_DELAY and LONG\_DELAY models differ only for the minimum value of the delay time $t_{\rm min}$ of $N$g mergers, which is 0.1 and 100 Myr, respectively. Finally, the BROAD\_VESC and NARROW\_VESC models differ from the fiducial case for the standard deviation of the log-normal distribution of $v_{\rm esc}$, which is 0.3 and 0.1 in the former and in the latter case.

\subsection{Merger rate}\label{sec:MRD}

We calculate the merger rate by assuming that each channel accounts for a fraction $f_{i}(t)$ of the star formation rate density at a given look-back time $t$ (where $i=$ NSC, GC, YSC or field).


In our fiducial model, we assume that $f_{\rm GC}(t)$ is given by
\begin{equation}\label{eq:frac_GC}
f_{\rm GC}(t)=f_{\rm max,GC}\,{}\exp{\left[-\frac{\left(t-t_{\rm GC}\right)^2}{2\,{}\sigma_t^2}\right]},
\end{equation}
where $f_{\rm max,GC}=0.1$, $t_{\rm GC}=11.8$ Gyr and $\sigma_t=2.5$ Gyr. The parameters $t_{\rm GC}$ and $\sigma_t$ are chosen based on the age distribution of Galactic GCs \citep{gratton1997,gratton2003,vandenberg2013}. 

NSCs  likely are the result of the dynamical assembly of GCs, which sink to the center of the galactic potential well by dynamical friction \citep{tremaine1975, capuzzo1993,capuzzo2008, antonini2012,antonini2013,arcasedda2014,arcasedda2015}, plus some contribution from in situ star formation \citep{mapelli2012}. Hence, we assume the same functional form for $f_{\rm NSC}(t)$, with a different normalization:

\begin{equation}
f_{\rm NSC}(t)=f_{\rm max,NSC}\,{}\exp{\left[-\frac{\left(t-t_{\rm GC}\right)^2}{2\,{}\sigma_t^2}\right]},
\end{equation}
where $f_{\rm max,NSC}=0.01$, while $t_{\rm GC}$ and $\sigma_t$ are the same as in eq.~\ref{eq:frac_GC}. The values of both $f_{\rm max,GC}$ and $f_{\rm max,NSC}$ are calibrated to give a mass budget of GCs and NSCs that matches the observed ones at low redshifts \citep{harris2013,neumayer2020}.

To keep the fiducial model as simple as possible, we assume
\begin{equation}
f_{\rm YSC}(t)=\min{\left\{0.3,\left[1-f_{\rm GC}(t)-f_{\rm NSC}(t)\right]\right\}}.
\end{equation}

Finally, we define $f_{\rm field}(t)=\max{\left\{0,\left[1-f_{\rm GC}(t)-f_{\rm NSC}(t)-f_{\rm YSC}(t)\right]\right\}}$. In the following sections, we briefly discuss the impact of changing the $f_i(t)$ parameters on the merger rate.

The total merger rate for each channel is then evaluated as
\begin{eqnarray}\label{eq:cosmorate}
   \mathcal{R}_i(z) = \frac{\rm d\quad{}\quad{}}{{\rm d}t(z)}\int_{z_{\rm max}}^{z}f_i(z')\,{}\psi(z')\,{}\frac{{\rm d}t(z')}{{\rm d}z'}\,{}{\rm d}z' \,{}   \int_{Z_{\rm min}(z')}^{Z_{\rm max}(z')}\eta{}(Z)\,{}\mathcal{F}(z',z, Z)\,{}{\rm d}Z,
\end{eqnarray}
where $t(z)$ is the look-back time at redshift $z$, $\psi(z')$ is the cosmic star formation rate density at redshift $z'$, $f_i(z')$ is the fraction of the total star formation rate that goes into channel $i=$~NSCs, GCs, YSCs or field at redshift $z'$, $Z_{\rm min}(z')$ and $Z_{\rm max}(z')$ are the minimum and maximum metallicity of stars formed at redshift $z'$, $\eta{}(Z)$ is the merger efficiency at metallicity $Z$, and $\mathcal{F}(z', z, Z)$ is the fraction of BBHs that form at redshift $z'$ from stars with metallicity $Z$ and merge at redshift $z$, normalized to all BBHs that form from stars with metallicity $Z$. To calculate the look-back time we take the cosmological parameters ($H_{0}$, $\Omega_{\rm M}$ and $\Omega_{\Lambda}$)  from \cite{planck2016}. The maximum considered redshift in equation~\ref{eq:cosmorate} is $z_{\rm max}=15$, which we assume to be the epoch of formation of the first stars. The merger efficiency $\eta{}(Z)$ is estimated as the number of BBHs that merge within a Hubble time in a coeval population of star with initial mass $M_\ast$ and metallicity $Z$, divided by $M_\ast$. We take the fitting formula for the star formation rate density $\psi{}(z')$ from  \cite{madau2017}:
\begin{equation}\label{eq:madau}
\psi{}(z)=0.01\,{}\frac{(1+z)^{2.6}}{1+[(1+z)/3.2]^{6.2}}~\text{M}_\odot\,{}\text{Mpc}^{-3}\,{}\text{yr}^{-1},
\end{equation}
 and we model the metallicity evolution as described in \cite{santoliquido2020}.  Equation~\ref{eq:madau} is based on data ranging from $z=0$ to $z\sim{10}$ \citep{madau2017}. 
 In future work, we will consider alternative options to model the very high redshift star formation, such as modelling population~III stars separately \cite{ng2021} and considering high-redshift long gamma-ray bursts as tracers of star formation (\cite{porciani2001}, but see \cite{palmerio2019} for some caveats).

\begin{figure}[H]
    \includegraphics[width = 12cm]{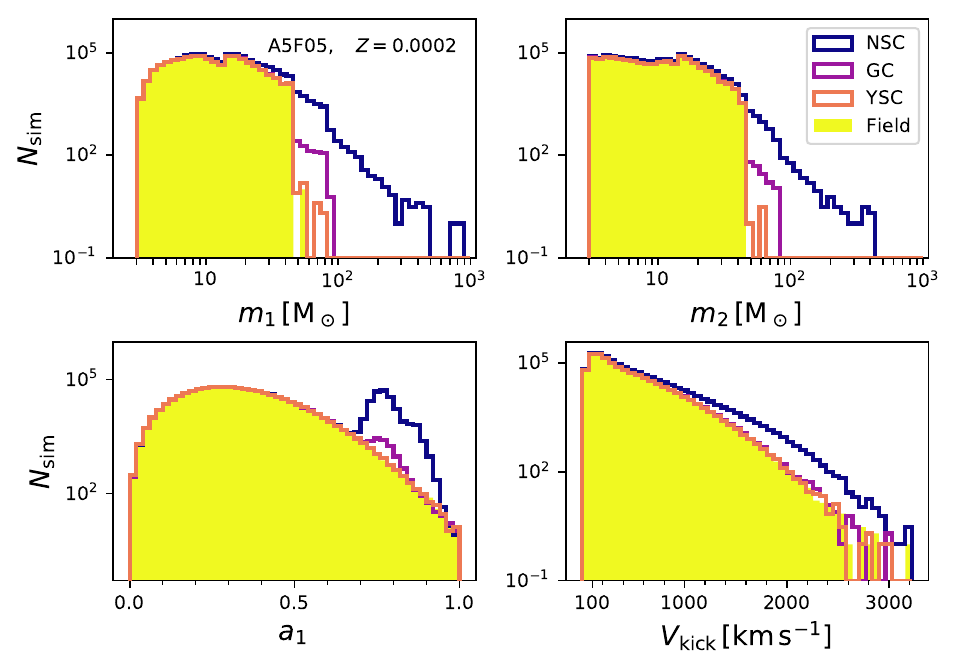}
\caption{Upper left (right): primary (secondary) mass distribution in the fiducial model A5F05 for $Z=0.0002$. Lower left (right): primary spin magnitude $a_1$ (relativistic kick velocity $v_{\rm kick}$) in the fiducial case for $Z=0.0002$. The distributions for each channel are drawn from an initial (i.e. zero-age main sequence) stellar population of $1.5\times{}10^{10}$ M$_\odot$, assuming a binary fraction $f_{\rm bin}=0.5$.  \label{fig:spin_Z0002}}
\end{figure}

   \section{Results} \label{sec:results}

   \subsection{Properties of hierarchical mergers}\label{sec:properties}

\begin{figure*}[]
  \includegraphics[width = 15cm]{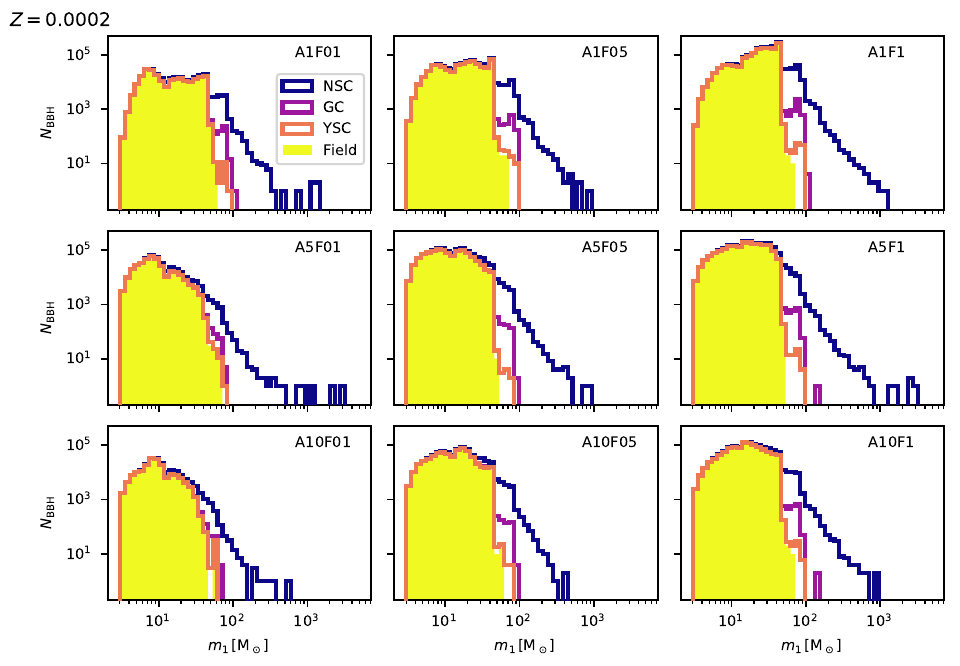}
\caption{From top to bottom and from left to right: distribution of primary BH masses ($m_1$), assuming progenitor metallicity $Z=0.0002$, in the cases A1F01 ($\alpha=1$, $f_{\rm MT}=0.1$), A1F05 ($\alpha=1$, $f_{\rm MT}=0.5$), A1F1 ($\alpha{}=1$, $f_{\rm MT}=1$), A5F01 ($\alpha=5$, $f_{\rm MT}=0.1$), A5F05 ($\alpha=5$, $f_{\rm MT}=0.5$, fiducial case), A5F1 ($\alpha=5$, $f_{\rm MT}=1$), A10F01 ($\alpha=10$, $f_{\rm MT}=0.1$), A10F05 ($\alpha=10$, $f_{\rm MT}=0.5$), and A10F1 ($\alpha=10$, $f_{\rm MT}=1$).  The distributions for each channel are drawn from an initial (i.e. zero-age main sequence) stellar population of $1.5\times{}10^{10}$ M$_\odot$, assuming a binary fraction $f_{\rm bin}=0.5$. \label{fig:AF_mass_Z0002}}
\end{figure*}
\begin{figure*}[]
  \includegraphics[width = 15cm]{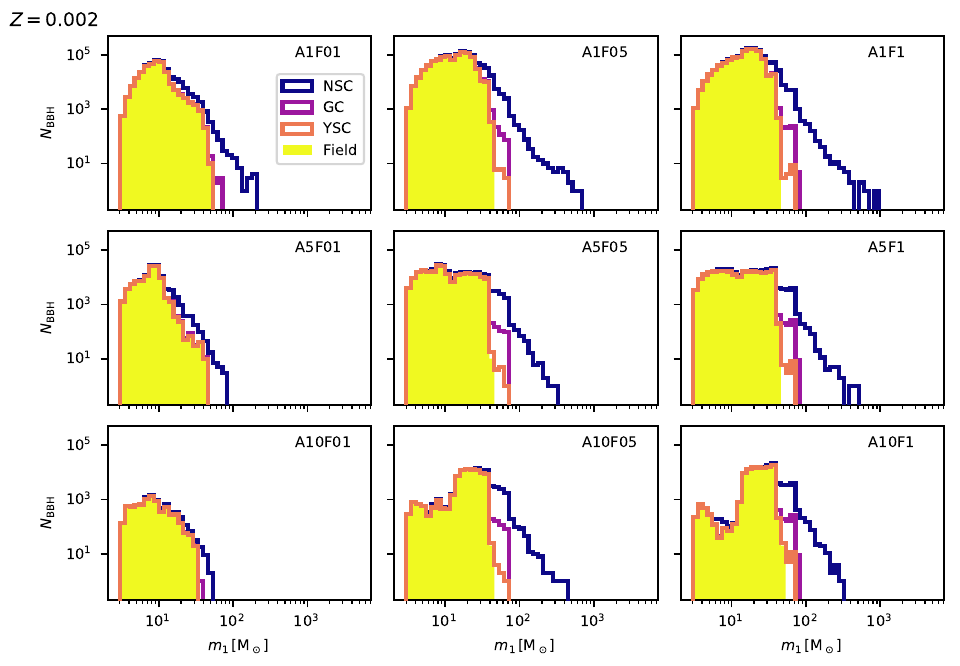}
\caption{Same as Figure~\ref{fig:AF_mass_Z0002} but for progenitor's metallicity $Z=0.002$.  \label{fig:AF_mass_Z002}}
\end{figure*}


Figure~\ref{fig:spin_Z0002} shows the mass of the primary BH ($m_1$), the mass of the secondary BH ($m_2$), the spin magnitude of the primary BH ($a_1$) and the kick velocity ($v_{\rm kick}$) in the fiducial model (A5F05) for $Z=0.0002$. The maximum primary and secondary mass strongly depend on the environment: we have 2g BBHs even in YSCs and GCs, but NSCs are more effective in producing hierarchical mergers, because of the larger value of $v_{\rm esc}$. In the fiducial model, the maximum primary mass is $\approx{}100$ M$_\odot$ in YSCs and GCs, while it is close to $\approx{}10^3$ M$_\odot$ in NSCs. The distribution of primary spin magnitudes shows a clear secondary peak at $a_1\approx{}0.7-0.8$ in both GCs and NSCs, corresponding to the typical values of $N$g merger remnants. The most common kick velocities are $v_{\rm kick}\sim{100-300}$ km s$^{-1}$, but larger kicks, up to $\sim{3000}$ km s$^{-1}$, are possible.

Figures~\ref{fig:AF_mass_Z0002} and \ref{fig:AF_mass_Z002} 
compare the primary mass distributions that we obtain by varying the values of $\alpha$ and $f_{\rm MT}$ in the first generation of BHs for metallicity $Z=0.0002$ and 0.002, 
respectively. By comparing Figures~\ref{fig:AF_mass_Z0002} and \ref{fig:AF_mass_Z002},  
it is apparent that both the BH mass distribution and the maximum BH mass strongly depend on progenitor's metallicity, even in hierarchical mergers. 
Also, the efficiency of common-envelope ejection $\alpha$ and the efficiency of mass accretion $f_{\rm MT}$ significantly affect the mass distribution of $N$g BHs. Hence, the mass distribution of 1g BBHs, which strongly depends on metallicity, has a crucial impact on the mass distribution of $N$g BHs.

\begin{figure*}[]
  \includegraphics[width = 15 cm]{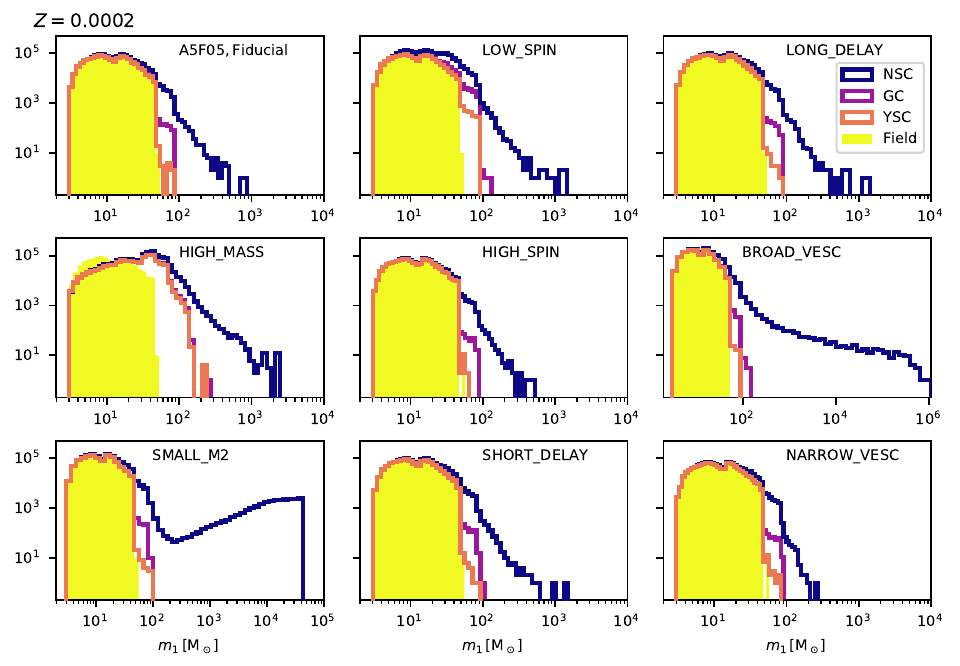}
\caption{From top to bottom and from left to right: distribution of primary BH masses ($m_1$), assuming progenitor metallicity $Z=0.0002$, in the cases A5F05 (fiducial), LOW\_SPIN, LONG\_DELAY, HIGH\_MASS, HIGH\_SPIN, BROAD\_VESC, SMALL\_M2,  SHORT\_DELAY, and NARROW\_VESC.  The distributions for each channel are drawn from an initial (i.e. zero-age main sequence) stellar population of $1.5\times{}10^{10}$ M$_\odot$, assuming a binary fraction $f_{\rm bin}=0.5$.  \label{fig:other_mass_Z0002}}   
\end{figure*}

In Figure~\ref{fig:other_mass_Z0002}, 
we fix $\alpha{}=5$, $f_{\rm MT}=0.5$ and $Z=0.0002$, and  consider the impact of the other main parameters of our model. Drawing the mass of 1g BBHs from the distribution of all 1g BHs 
(including single BHs) shifts the entire distribution of dynamical mergers to higher masses. In the model HIGH\_MASS, the most common primary mass of dynamical BBHs  is $\sim{}30-50$ M$_\odot$, while the primary masses of field BBHs peak at $\sim{}10$ M$_\odot$. The reason is that {\sc mobse} allows the formation of BHs with mass up to $\sim{}65$ M$_\odot$, but small BHs merge more efficiently than  massive BHs, because of the interplay between stellar radii, mass transfer and common envelope evolution. If we randomly pair single BHs from {\sc mobse} data, this effect disappears. Hence, the HIGH\_MASS model is realistic if dynamical encounters are very effective, and all  BBHs in star clusters form from dynamical exchanges. Based on direct $N-$body simulations coupled with {\sc mobse}, \cite{dicarlo2020} have shown that BBHs in YSCs behave in an intermediate way between the HIGH\_MASS model and our fiducial model. 
In the HIGH\_MASS model, $N$g BHs in both GCs and YSCs can reach masses $m_1\sim{}200$ M$_\odot$, while the maximum mass of $N$g BHs in NSCs is $\sim{}2000$ M$_\odot$.

In the SMALL\_M2 model, we  draw the secondary mass from the distribution of 1g BHs, which is equivalent to assuming that only $N$g--1g mergers are possible. 
Hence, this model differs from the others because of the smaller values of $q=m_2/m_1$ in hierarchical mergers. We observe a peculiar trend of $m_1$ in the NSC case: values of $m_1\sim{}10^4$ M$_\odot$ are about one order of magnitude more common than $m_1\sim{}200$ M$_\odot$. The reason is that relativistic kicks get smaller and smaller if $q$ tends to zero. Hence, the maximum mass of the primary BH in this model is set by the number of hierarchical mergers that happen within a Hubble time, rather than by the relativistic kicks.

If we compare the LOW\_SPIN ($\sigma_a=0.01$) and  HIGH\_SPIN model ($\sigma_a=0.4$), we see that high-mass BHs are more and more suppressed if the spin distribution moves to higher values, because relativistic kicks get stronger. In contrast, we find only a mild difference between the SHORT\_DELAY and  LONG\_DELAY model, in which we change the minimum delay time $t_{\rm min}$.

Finally, the escape velocity has a large impact on hierarchical BH masses, especially for NSCs. BHs with mass up to $\sim{}10^6$ M$_\odot$ form if $\sigma_{\rm v}=0.3$ (BROAD\_VESC), two orders of magnitude more  than if $\sigma_{\rm v}=0.2$ (fiducial case, A5F05) and three orders of magnitude more than if $\sigma_{\rm v}=0.1$ (NARROW\_VESC). This result is consistent with both \cite{antonini2019} and \cite{fragione2020}, who report that the maximum BH mass approaches $10^6$ M$_\odot$ if $v_{\rm esc}\ge{}300$ km s$^{-1}$ are considered. This might be a key ingredient to understand the formation of super-massive BHs and the connection between the mass of the central BH and its parent galaxy mass/central velocity dispersion \citep{ferrarese2000,gebhardt2000,graham2009,neumayer2020}.

\begin{specialtable}[] 
\caption{Values of $f_{>1g}$, $f_{\rm PISN}$ and $f_{\rm IMBH}$ for  different runs.\label{tab:table2}}
\footnotesize 
\begin{tabular}{ccccc}
  \toprule
  \textbf{Run Name} & \textbf{Star cluster} & \textbf{$f_{>1g}$} & \textbf{$f_{\rm PISN}$}& \textbf{$f_{\rm IMBH}$}\\
\midrule
Fiducial, A5F05 & NSC & $0.16,\,{}0.15,\,{}0.13$ &  $0.007,\,{}0.009,\,{}0$ &  $5\times{}10^{-4},\,{}5\times{}10^{-4},\,{}0$ \\
 & GC & $0.006,\,{}0.007,\,{}0.005$ &  $3\times{}10^{-4},\,{}5\times{}10^{-4},\,{}0$ &  $0,\,{}0,\,{}0$ \\
 & YSC & $10^{-4},\,{}2\times{}10^{-4},\,{}0$ &  $5\times{}10^{-6},\,{}2\times{}10^{-5},\,{}0$ &  $0,\,{}0,\,{}0$ \vspace{0.1cm}\\
A5F01 & NSC & $0.14,\,{}0.11,\,{}0.13$ & $0.003,\,{}7\times{}10^{-5},\,{}0$ & $2\times{}10^{-4},\,{}0,\,{}0$\\
 & GC & $0.005,\,{}0.002,\,{}0$ & $2\times{}10^{-4},\,{}0,\,{}0$ & $0,\,{}0,\,{}0$\\
 & YSC & $10^{-4},\,{}5\times{}10^{-5},\,{}0$ & $10^{-5},\,{}0,\,{}0$ & $0,\,{}0,\,{}0$ \vspace{0.1cm}\\
A5F1 & NSC & $0.14,\,{}0.16,\,{}0.15$ & $0.014,\,{}0.020,\,{}0$ & $8\times{}10^{-4},\,{}8\times{}10^{-4},\,{}0$\\
 & GC & $0.004,\,{}0.008,\,{}0.004$ & $7\times{}10^{-4},\,{}0.0013,\,{}0$ & $5\times{}10^{-7},\,{}0,\,{}0$\\
 & YSC & $9\times{}10^{-5},\,{}3\times{}10^{-4},\,{}0.004$ & $2\times{}10^{-5},\,{}4\times{}10^{-5},\,{}0$ & $0,\,{}0,\,{}0$ \vspace{0.1cm}\\
HIGH\_MASS & NSC & $0.26,\,{}0.25,\,{}0.15$ & $0.075,\,{}0.055,\,{}0$ & $0.016,\,{}0.011,\,{}0$\\
 & GC & $0.009,\,{}0.007,\,{}0.005$ & $0.004,\,{}0.003,\,{}0$ & $6\times{}10^{-4},\,{}6\times{}10^{-4},\,{}0$\\
& YSC & $9\times{}10^{-5},\,{}8\times{}10^{-5},\,{}0$ & $4\times{}10^{-5},\,{}3\times{}10^{-5},\,{}0$ & $3\times{}10^{-5},\,{}3\times{}10^{-5},\,{}0$ \vspace{0.1cm}\\
SMALL\_M2 & NSC & $0.18,\,{}0.19,\,{}0.14$ & $0.009,\,{}0.014,\,{}0$ & $0.019,\,{}0.038,\,{}0$\\
 & GC & $0.006,\,{}0.007,\,{}0.003$ & $3\times{}10^{-4},\,{}4\times{}10^{-4},\,{}0$ & $0,\,{}0,\,{}0$\\
& YSC & $10^{-4},\,{}2\times{}10^{-4},\,{}0.0026$ & $10^{-5},\,{}10^{-5},\,{}0$ & $0,\,{}0,\,{}0$ \vspace{0.1cm}\\
LOW\_SPIN & NSC & $0.45,\,{}0.43,\,{}0.37$ & $0.021,\,{}0.029,\,{}0$ & $9\times{}10^{-4},\,{}0.0011,\,{}0$\\
 & GC & $0.11,\,{}0.15,\,{}0.11$ & $0.007,\,{}0.012,\,{}0$ & $2\times{}10^{-6},\,{}4\times{}10^{-6},\,{}0$\\
& YSC & $0.013,\,{}0.020,\,{}0.013$ & $0.001,\,{}8\times{}10^{-4},\,{}0$ & $0,\,{}0,\,{}0$ \vspace{0.1cm}\\
HIGH\_SPIN & NSC & $0.08,\,{}0.08,\,{}0.09$ & $0.004,\,{}0.004,\,{}0$ & $2\times{}10^{-4},\,{}3\times{}10^{-4},\,{}0$\\
 & GC & $0.002,\,{}0.003,\,{}0.003$ & $10^{-4},\,{}2\times{}10^{-4},\,{}0$ & $0,\,{}0,\,{}0$\\
& YSC & $4\times{}10^{-5},\,{}4\times{}10^{-5},\,{}0$ & $9\times{}10^{-7},\,{}0,\,{}0$ & $0,\,{}0,\,{}0$ \vspace{0.1cm}\\
SHORT\_DELAY & NSC & $0.16,\,{}0.15,\,{}0.12$ & $0.007,\,{}0.010,\,{}0$ & $5\times{}10^{-4},\,{}6\times{}10^{-4},\,{}0$\\
 & GC & $0.006,\,{}0.007,\,{}0.005$ & $3\times{}10^{-4},\,{}5\times{}10^{-4},\,{}0$ & $9\times{}10^{-7},\,{}0,\,{}0$\\
& YSC & $10^{-4},\,{}2\times{}10^{-4},\,{}0$ & $6\times{}10^{-6},\,{}9\times{}10^{-6},\,{}0$ & $0,\,{}0,\,{}0$ \vspace{0.1cm}\\
LONG\_DELAY & NSC & $0.16,\,{}0.15,\,{}0.14$ & $0.007,\,{}0.010,\,{}0$ & $4\times{}10^{-4},\,{}6\times{}10^{-4},\,{}0$\\
 & GC & $0.006,\,{}0.007,\,{}0.003$ & $3\times{}10^{-4},\,{}5\times{}10^{-4},\,{}0$ & $0,\,{}0,\,{}0$\\
& YSC & $10^{-4},\,{}2\times{}10^{-4},\,{}0$ & $7\times{}10^{-6},\,{}10^{-5},\,{}0$ & $0,\,{}0,\,{}0$ \vspace{0.1cm}\\
BROAD\_VESC & NSC & $0.20,\,{}0.20,\,{}0.16$ & $0.012,\,{}0.015,\,{}0$ & $0.004,\,{}0.005,\,{}0$\\
 & GC & $0.012,\,{}0.013,\,{}0.003$ & $6\times{}10^{-4},\,{}9\times{}10^{-4},\,{}0$ & $4\times{}10^{-6},\,{}8\times{}10^{-6},\,{}0$\\
& YSC & $4\times{}10^{-4},\,{}5\times{}10^{-4},\,{}0$ & $3\times{}10^{-5},\,{}3\times{}10^{-5},\,{}0$ & $0,\,{}0,\,{}0$ \vspace{0.1cm}\\
NARROW\_VESC & NSC & $0.14,\,{}0.13,\,{}0.08$ & $0.006,\,{}0.007,\,{}0$ & $10^{-4},\,{}\times{}10^{-4},\,{}0$\\
 & GC & $0.003,\,{}0.005,\,{}0$ & $2\times{}10^{-4},\,{}3\times{}10^{-4},\,{}0$ & $0,\,{}0,\,{}0$\\
& YSC & $8\times{}10^{-5},\,{}10^{-4},\,{}0$ & $7\times{}10^{-6},\,{}4\times{}10^{-6},\,{}0$ & $0,\,{}0,\,{}0$ \vspace{0.1cm}\\
\bottomrule
\end{tabular}
\flushleft{Column 1: Name of the model.  Column 2: star cluster type (NSC, GC or YSC). Column 3: fraction of $N$g BBHs $f_{>1g}$. The three values reported in each line refer to $Z=0.0002$, 0.002 and 0.02. Column 4: fraction of BBHs with primary mass in the pair instability gap $f_{\rm PISN}$. The three values reported in each line refer to $Z=0.0002$, 0.002 and 0.02. Column 5: fraction of IMBH mergers $f_{\rm IMBH}$. The three values reported in each line refer to $Z=0.0002$, 0.002 and 0.02.}
\end{specialtable}

\begin{figure}[]
   \includegraphics[width = 10.5cm ]{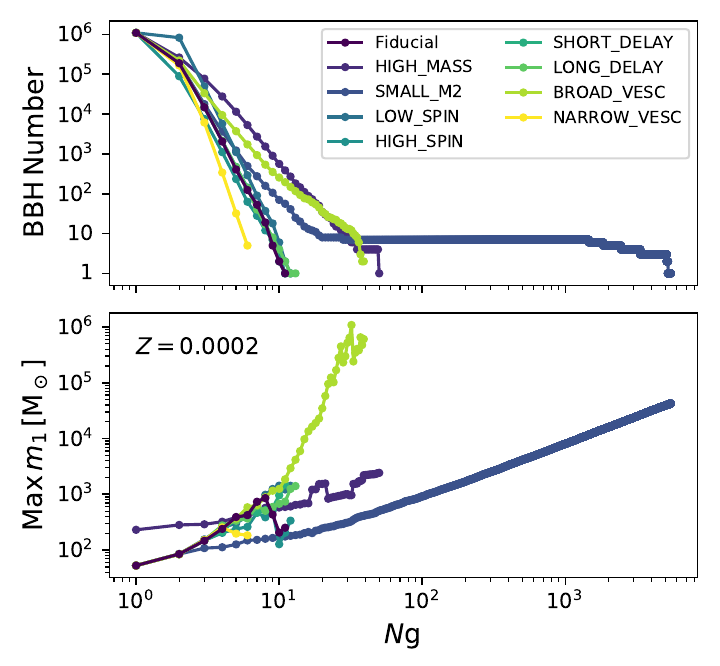} 
\caption{Top: number of BBHs in each generation as a function of the generation number $N$g for hierarchical BBHs in NSCs with metallicity $Z=0.0002$. Bottom: maximum primary BH mass  in each generation as a function of the generation number $N$g for hierarchical BBHs in NSCs with metallicity $Z=0.0002$. We show models A5F05 (fiducial), HIGH\_MASS, SMALL\_M2, LOW\_SPIN,  HIGH\_SPIN,   SHORT\_DELAY, LONG\_DELAY,  BROAD\_VESC, and NARROW\_VESC. \label{fig:ngen_Z0002}}
\end{figure}

Figure~\ref{fig:ngen_Z0002} shows the number of BBHs we simulated per each generation $N$g in the case of $Z=0.0002$ and the maximum primary mass in each generation. We only show NSCs because BHs in GCs and YSCs do not exceed the 5th and  3rd generation, respectively. The maximum number of generations in NSCs ranges from a few to a few thousands. In the fiducial case and in most of the other simulations, the maximum number of generations is $N\sim{}10$. Only in three cases we obtain a significantly larger number of generations, namely the BROAD\_VESC model ($\approx{}40$ generations), the HIGH\_MASS model ($\approx{}50$ generations) and the SMALL\_M2 model ($\approx{}5000$ generations). The SMALL\_M2 case outnumbers all the other models for the number of generations, because of the strong dependence of $v_{\rm kick}$ on $q$. However, even in this extreme case, the number of $N$g mergers with $N\ge{}20$ is $\sim{}10^5$ times lower than the number of mergers in the first generation.

\begin{figure}[]
  \includegraphics[width = 13cm]{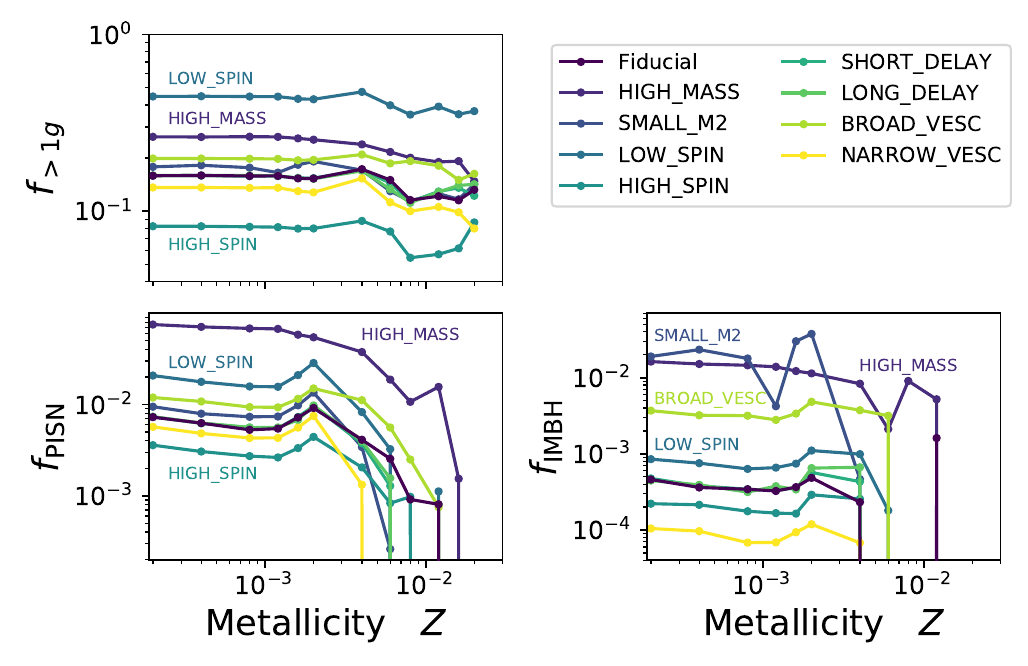}
\caption{Top: $f_{>1g}$ is the number of $N$g BBH mergers with $N>1$ divided by the total number of BBH mergers. Bottom left: $f_{\rm PISN}$ is the number of  $N$g BBH mergers with a primary mass in the pair instability mass gap ($m_1\in{}[60,120]\,{}{\rm M}_\odot$), divided by the total  number of BBH mergers. Bottom right: $f_{\rm IMBH}$ is the number of  $N$g BBH mergers with a primary mass in the IMBH regime ($m_1\ge{}100$ M$_\odot$), divided by the total number of BBH mergers.  $f_{>1g}$, $f_{\rm PISN}$ and $f_{\rm IMBH}$ refer to NSCs only and are shown as a function of the metallicity $Z$. We show models A5F05 (fiducial),  HIGH\_MASS, SMALL\_M2, LOW\_SPIN,  HIGH\_SPIN,   SHORT\_DELAY, LONG\_DELAY,  BROAD\_VESC, and NARROW\_VESC.\label{fig:ngen_PISN}}
\end{figure}

The  upper panel of Fig.~\ref{fig:ngen_PISN} shows the fraction of $N$g BBH mergers with $N>1$ with respect to all BBH mergers, defined as
  $f_{>1g}=(N_{2g}+N_{3g}+..+N_{Ng})/N_{\rm BBH}$,
where $N_{1g}$, $N_{2g}$, $N_{3g}$,..,$N_{Ng}$ is the number of 1g, 2g, 3g,..,$N$g BBH mergers  and $N_{\rm BBH}$ is the total number of BBH mergers summing up all possible generations including the first one. In this figure, $f_{>1g}$ is only shown  for NSCs. In the fiducial model and in NSCs, $N$g BBH mergers with $N>1$ are about 16\% of all the BBH mergers, with a small dependence on metallicity. For other models, the percentage of $N$g BBHs can be as low as $\sim{}8$\% (HIGH\_SPIN case) or as high as $\sim{}40-50$\% (LOW\_SPIN case). For GCs and YSCs these percentages should be lowered by a factor of $\sim{}30$ and $\sim{}10^3$, respectively. 
Table~\ref{tab:table2} reports the values of $f_{>1g}$ in detail.

\subsection{BHs in the mass gap and IMBHs}


Hierarchical mergers could be responsible for the formation of BHs with mass in the pair instability mass gap ($\sim{}60-120$ M$_\odot$) or even in the IMBH regime ($>100$ M$_\odot$). The bottom left panel of Figure~\ref{fig:ngen_PISN} shows $f_{\rm PISN}$ defined as $f_{\rm PISN}=N_{\rm PISN}/N_{\rm BBH}$, 
where $N_{\rm PISN}$ is the number of BBH mergers with primary mass in the pair instability mass gap, while $N_{\rm BBH}$ is the number of all BBH mergers. In our fiducial model and in NSCs, $\sim{}0.7$\% of all BBH mergers contain at least one BH in the pair instability mass gap at the lowest metallicity ($Z=0.0002$). This percentage decreases as metallicity increases and drops to zero at $Z\ge{}0.012$. The other models follow the same trend with metallicity. The HIGH\_MASS model is the one with the largest value of $f_{\rm PISN}$: in this case, up to 7.5\% of all the BBH mergers  contain at least one BH in the pair instability mass gap at the lowest metallicity ($Z=0.0002$). These percentages should be lowered by a factor of $\gtrsim{}10$ in GCs and by a factor of $\sim{}10^3$ in YSCs.

In the bottom right panel of Figure~\ref{fig:ngen_PISN}, we show the fraction of IMBH mergers $f_{\rm IMBH}$, defined as 
$f_{\rm IMBH}=N_{\rm IMBH}/N_{\rm BBH}$, where $N_{\rm IMBH}$ is the number of BBH mergers with primary mass $m_1>10^2$ M$_\odot$. The fraction of IMBH mergers follows the same trend with metallicity as $f_{\rm PISN}$: it is higher at lower $Z$ and drops to zero at $Z\ge{}4\times{}10^{-3}$. In the fiducial model, $f_{\rm IMBH}\sim{}5\times{}10^{-4}$ at $Z=0.0002$ in NSCs. We find no IMBHs in GCs and YSCs in the fiducial case. The fraction of IMBH mergers is maximum in the SMALL\_M2 simulation, where  $f_{\rm IMBH}\sim{}2\times{}10^{-2}$ at $Z=0.0002$. Moreover,  $f_{\rm IMBH}\sim{}6\times{}10^{-4}$ and $\sim{}3\times{}10^{-5}$ at $Z=0.0002$ in the HIGH\_MASS case for GCs and YSCs, respectively. 
Table~\ref{tab:table2} reports the values of $f_{\rm PISN}$ and $f_{\rm IMBH}$ in detail.

\subsection{Merger rates}
\begin{figure*}[]
  \includegraphics[width = 17cm]{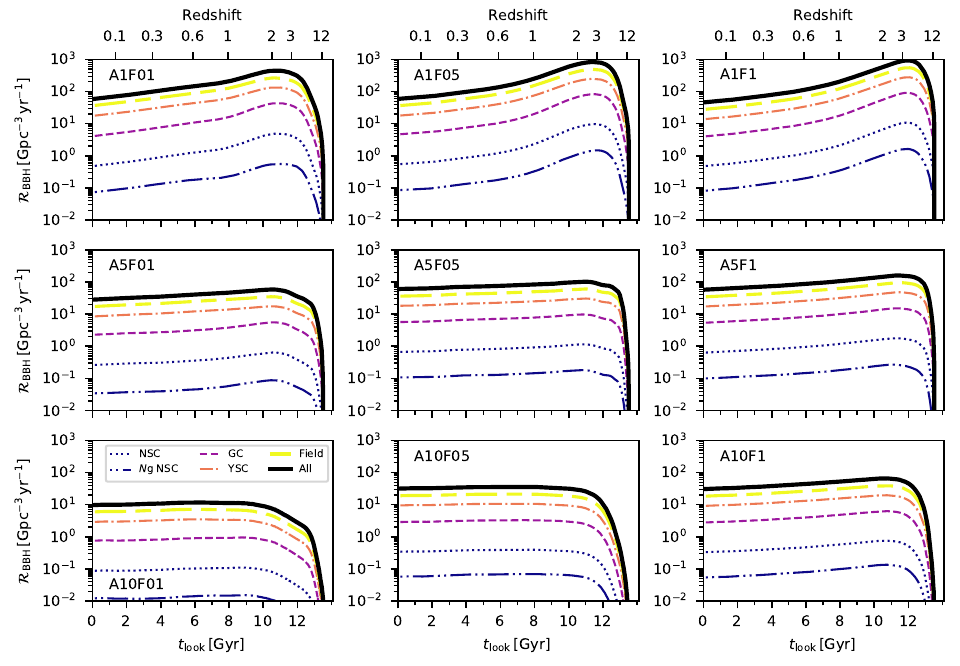}
  \caption{From top to bottom and from left to right: BBH merger rate density in the comoving frame as a function of look-back time (bottom $x-$axis) and redshift (top $x-$axis) for the simulations A1F01, A1F05, A1F1, A5F01, A5F05, A5F1, A10F01, A10F05, and A10F1. Black thick line: total merger rate density; yellow long--short dashed line: BBH merger rate density from field binaries; pink dot-dashed line: BBH  merger rate density from YSCs; violet dashed line:  BBH merger rate density from GCs; blue dotted line: BBH merger rate density from NSCs; blue dot-dot-dashed line: BBH merger rate density from NSCs if we consider only $N$g BBHs.\label{fig:MRD}}
\end{figure*}

\begin{figure*}[]
  \includegraphics[width = 17cm]{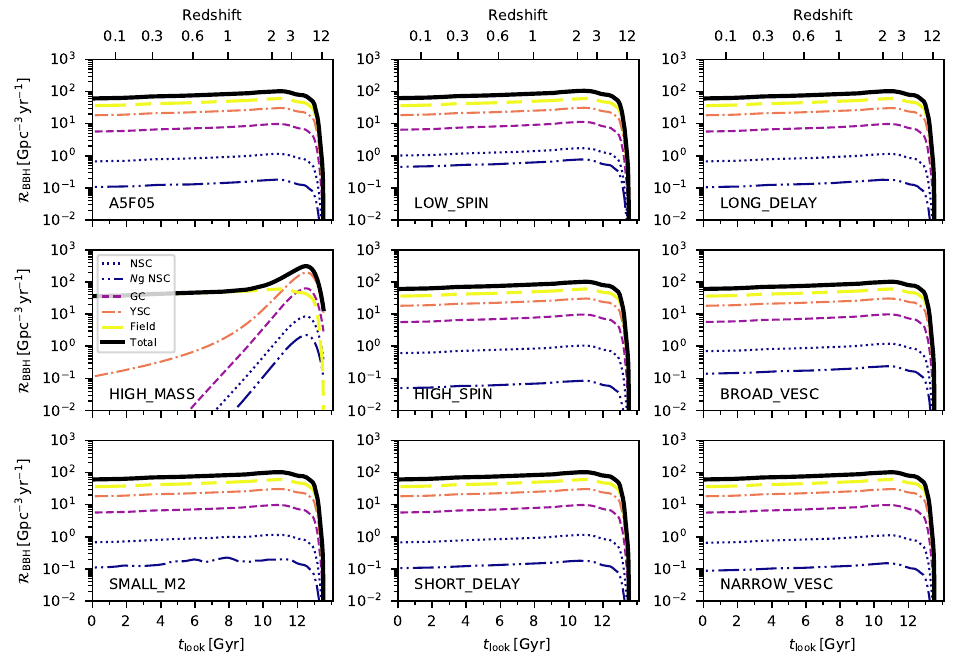}
\caption{Same as Figure~\ref{fig:MRD} but for the simulations A5F05 (fiducial), LOW\_SPIN, LONG\_DELAY, HIGH\_MASS, HIGH\_SPIN, BROAD\_VESC, SMALL\_M2,  SHORT\_DELAY, and NARROW\_VESC.  \label{fig:MRD_other}}
\end{figure*}

Figures~\ref{fig:MRD} and \ref{fig:MRD_other} show the merger rate density evolution for all our models, calculated as detailed in Section~\ref{sec:MRD}. The contribution of each channel to the total merger rate density is 
set by the value of $f_i(z)$, because  hierarchical mergers are only a small fraction of the total BBH mergers (Figure~\ref{fig:ngen_PISN}). Since $f_i(z)$ is highly uncertain, the relative importance of different channels in Figures~\ref{fig:MRD} and \ref{fig:MRD_other} can change wildly and is only indicative. The uncertainty is particularly large for field and YSCs.

Models with $\alpha{}=1$ have a higher merger rate than models with $\alpha=5,\,{}10$. 
The merger rate evolution of dynamical BBHs in the HIGH\_MASS case is remarkably different from the other cases. The reason is our choice of the delay time distribution of 1g BBHs ($dN/dt\propto{}t^{-1}$), which does not take into account a possible dependence of $t_{\rm delay}$ on the mass and other properties of BBHs. 
In particular, the delay time distribution obtained with {\sc mobse} tends to deviate from the $dN/dt\propto{}t^{-1}$ trend when $t_{\rm delay}<1$ Gyr. Hence, dynamical BBHs in the HIGH\_MASS case have shorter delay times than the fiducial case.

Figures~\ref{fig:MRD} and \ref{fig:MRD_other} also show the BBH merger rate density we obtain if we consider only $N$g BBHs  in NSCs. In the local Universe, the merger rate density of $N$g BBHs in NSCs ranges from $\sim{}10^{-2}$ to $\sim{}0.2$ Gpc$^{-3}$ yr$^{-1}$. For GCs and YSCs we obtain lower values because, even if these star clusters are likely more common than NSCs, the occurrence of $N$g BBH mergers in GCs and YSCs is lower than in NSCs (e.g. Section~\ref{sec:properties}).

\subsection{Mass distribution at different redshifts}


\begin{figure*}[]
  \includegraphics[width = 17cm]{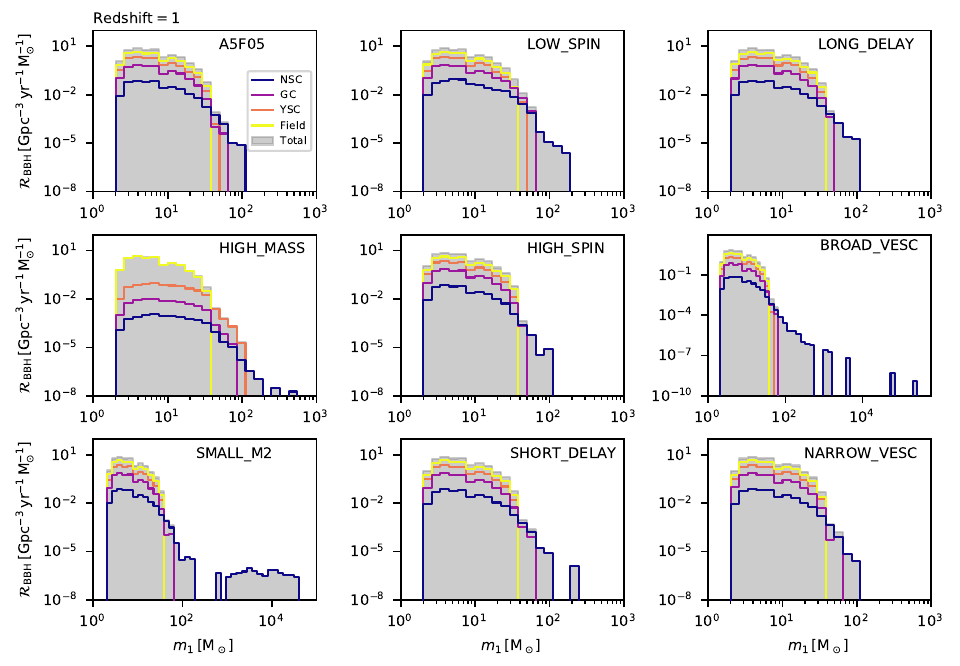}
  \caption{ From top to bottom and from left to right: merger rate density per unit primary mass  as a function of BBH mass at redshift $z=1$ for the simulations  A5F05 (fiducial), LOW\_SPIN, LONG\_DELAY, HIGH\_MASS, HIGH\_SPIN, BROAD\_VESC, SMALL\_M2,  SHORT\_DELAY, and  NARROW\_VESC. Grey histogram: sum of all formation channels; yellow: field binaries; pink: YSCs; violet: GCs; blue: NSCs. \label{fig:mass_z1}}
\end{figure*}



\begin{figure*}[]
  \includegraphics[width = 17cm]{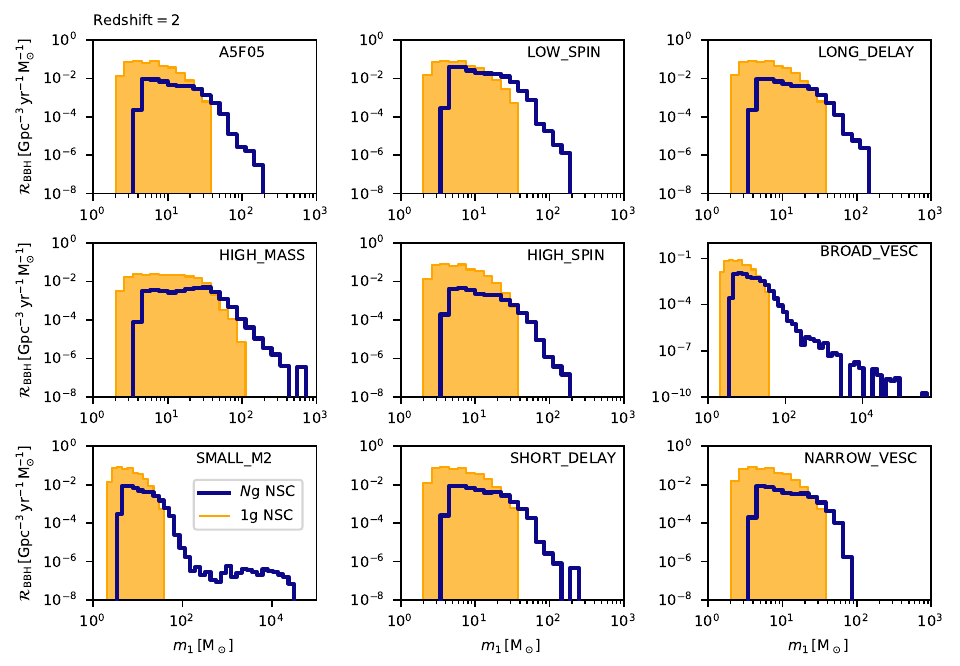}
  \caption{Merger rate density per unit primary mass  as a function of BBH mass at redshift $z=2$ for BBHs in NSCs. Filled orange histogram: 1g BBHs; blue histogram: $N$g BBHs. From top to bottom and from left to right: simulations  A5F05 (fiducial), LOW\_SPIN, LONG\_DELAY, HIGH\_MASS, HIGH\_SPIN, BROAD\_VESC, SMALL\_M2,  SHORT\_DELAY, and NARROW\_VESC. \label{fig:mass_z2}}
\end{figure*}


Figure~\ref{fig:mass_z1} 
shows the total mass distribution of primary BHs in the source frame at redshift $z=1$.
NSCs are responsible for the high mass tail ($m_1\gtrsim{}100$ M$_\odot$) at all redshifts and in all models. We show only the distribution at $z=1$, because we do not see significant changes of the mass distribution with redshift in all cases but the HIGH\_MASS model. In this case, the importance of dynamical BBHs drops at redshift zero because of the different delay time distributions (Figure~\ref{fig:MRD_other}). 

Figure~\ref{fig:mass_z2} shows the mass distribution of primary BHs at redshift $z=2$ for NSCs only. We separate 1g BBHs from $N$g BBHs with $N>1$. The maximum mass of 1g BBHs extends up to $\sim{}40$ M$_\odot$ in all simulations but the HIGH\_MASS case. In the HIGH\_MASS case, 1g BHs with mass up to $\sim{}100$ M$_\odot$ are possible, because this model includes BHs that form with mass in the pair instability gap from the merger of massive stars  \citep{dicarlo2020} and acquire companions by dynamical exchanges.

The mass of $N$g BHs extends up to $\sim{}100-200$ M$_\odot$ in most models, with the exception of the following runs. In the HIGH\_MASS case, we find primary BHs with mass up to $\sim{}600-10^3$ M$_\odot$. In the SMALL\_M2 case, the most massive BH reaches $\sim{}5\times{}10^4$ M$_\odot$. Finally, this realization of the BROAD\_VESC model produces one single BH with mass $\sim{}4.3\times{}10^5$ M$_\odot$. To obtain the shown distributions, we started from catalogs of $\ge{}10^6$ BBHs. 
Figure~\ref{fig:mass_z2} confirms that the distribution of $N$g BBHs strongly depends not only on the properties of the environment (e.g. $v_{\rm esc}$) but also on the mass distribution of 1g BHs.

\begin{figure*}[]
  \includegraphics[width = 17cm]{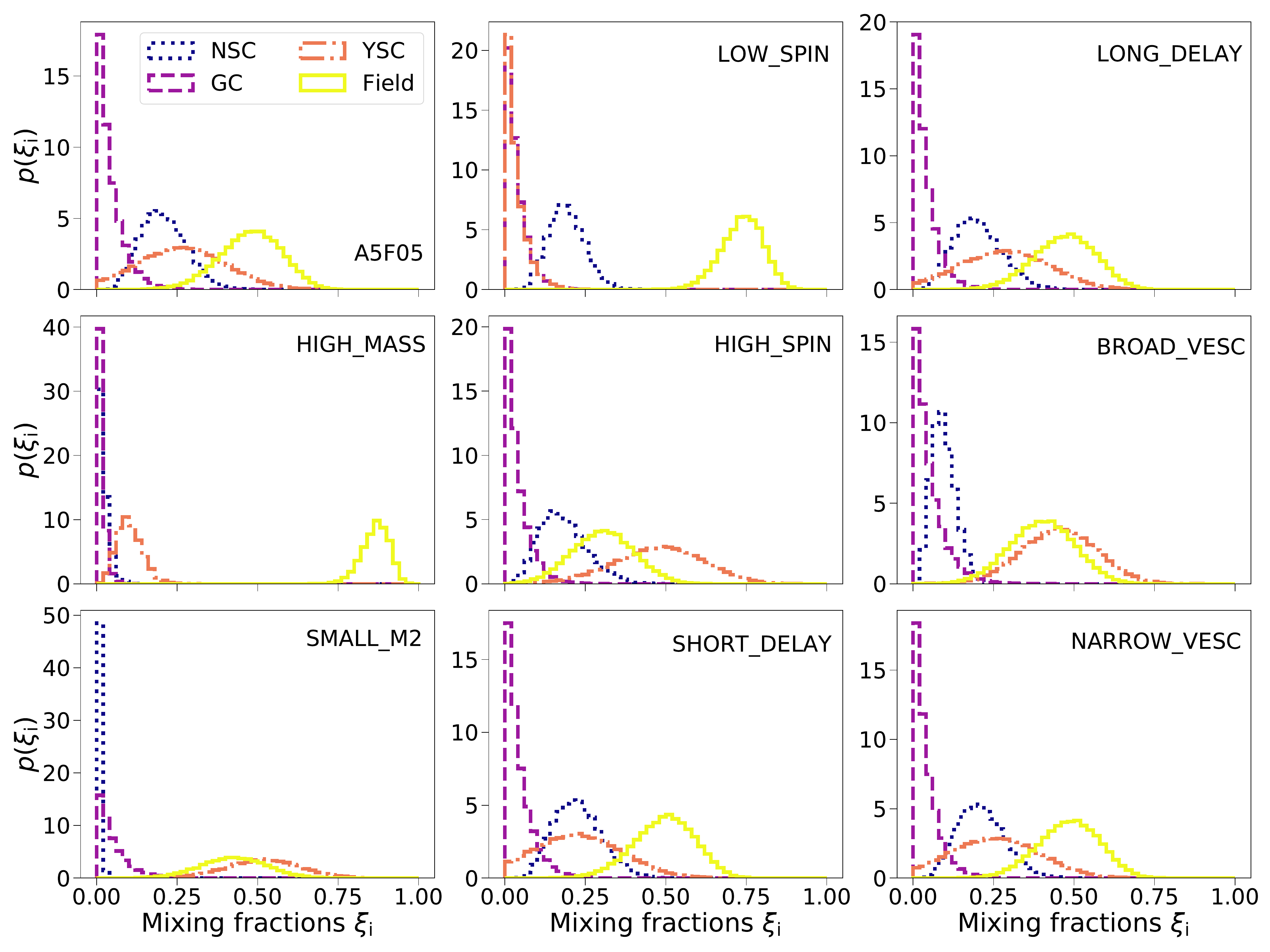}
\caption{Posterior probability distribution of the mixing fractions $\xi{}_{\rm i}$ for our multi-channel analysis. The mixing fractions $\xi{}_{\rm i}$, where ${\rm i}=1,$ 2, 3 and 4 for Field, YSCs, GCs and NSCs, are defined in eq.~\ref{eq:mixfrac}. From top to bottom and from left to right: simulations A5F05 (fiducial), LOW\_SPIN, LONG\_DELAY, HIGH\_MASS, HIGH\_SPIN, BROAD\_VESC, SMALL\_M2,  SHORT\_DELAY, and NARROW\_VESC. \label{fig:mix_frac}}
\end{figure*}

\subsection{Comparison with BBHs in GWTC-2}

To compare our models against GW events in the first (O1), second (O2) and in the first part of the third observing run (O3a) of the LIGO--Virgo collaboration  (hereafter, the GWTC-2 catalog \cite{abbottO3a}), we use a hierarchical Bayesian approach. In this framework, the posterior for a set of data $\lbrace h \rbrace^{k}$ observed during an observation time $T_{obs}$ and a model parametrized by $\lambda$ is well described by an in-homogeneous  Poisson process \citep{loredo2004,mandel2018}
 
\begin{eqnarray}
p(\lambda{}, N_\lambda | \lbrace h \rbrace^{k}) \sim \text{e}^{-\mu_{\lambda}}\,{}  \pi(\lambda{}, N_\lambda{}) \prod_{k=1}^{N_{obs}}  N_{\lambda} \int_{\theta} \mathcal{L}^{k}(\lbrace h\rbrace^k | \theta) \,{}p(\theta | \lambda )\,{}{\rm d}\theta{},\nonumber{}\\
\label{eq:post_hier_model}
\end{eqnarray}
where $\theta$ are the GW parameters, $N_{\lambda}$ is the number of events predicted by the astrophysical model, $\mu_{\lambda}$ is the predicted number of detections associated with the model and  GW detector, $\pi{}(\lambda{},N_\lambda{})$ is the prior distribution on $\lambda$ and $N_\lambda$, and $\mathcal{L}^{k}(\lbrace h\rbrace^k | \theta)$ is the likelihood of the $k-$th detection. 

The predicted number of detections is given by $\mu{}(\lambda{})=N_\lambda\,{}\beta{}(\lambda{})$, where $\beta{}(\lambda{})=\int_\theta p(\theta{}|\lambda{})$ $\,{}p_{\rm det}(\theta{})\,{}{\rm d}\theta$ is the detection efficiency of the model; $p_{\rm det}(\theta{})$ is the probability of detecting a source with parameters $\theta$ and can be inferred by computing the optimal signal-to-noise ratio and comparing it to a detection threshold \citep{bouffanais2021}. The values for the event's log-likelihood are derived from the posterior and prior samples released by the LIGO--Virgo collaboration, such that the integral in the above equation is approximated with a Monte Carlo approach as 
\begin{equation}
\int_{\theta}\mathcal{L}^{k}(\lbrace h\rbrace^k | \theta) \,{}p(\theta | \lambda )\,{}{\rm d}\theta{}\sim{}\frac{1}{N_s^k}\,{}\sum_{i=1}^{N_s^k}\frac{p(\theta^k_i | \lambda{})}{\pi^k(\theta_i^k)},
\end{equation}
where $\theta_i^k$ is the $i-$th posterior sample for the $k-$th detection and $N_s^k$ is the total number of posterior samples for the $k-$th detection. To compute the prior term in the denominator, we use a Gaussian kernel density estimation.

In practice, for each model, we generate a catalog of a fixed number of sources (fixed to $50 000$ sources), such that the sources are distributed according to the merger rate density of the model. 
Each entry of the catalog is represented by a set of parameters $\theta=\{\mathcal{M}_{c},\,{}q\,{},\chi_{\rm eff},\,{}z\}$, where $\mathcal{M}_{c}$ is the chirp mass of the source, $q$ the mass ratio, $\chi_{\rm eff}$ is the effective spin and $z$ the redshift, that was set to take values between 0 and 2. 
More details on 
this procedure are described in \cite{mandel2018} and \cite{bouffanais2021}.

In our analysis, our model distribution is the sum of the contributions from multiple channels (isolated BBHs, dynamical BBHs in YSCs, GCs and NSCs) weighted by mixing fraction hyper-parameters as
\begin{equation}\label{eq:mixfrac}
p(\theta{}|\xi{}_1,\,{}\xi{}_2,\,{},\xi{}_3,\,{}\xi{}_4,\lambda{})=\xi{}_1\,{}p(\theta{}|{\rm Field},\,{}\lambda{})+\xi{}_2\,{}p(\theta{}|{\rm YSC},\,{}\lambda{})+\xi{}_3\,{}p(\theta{}|{\rm GC},\,{}\lambda{})+\xi{}_4\,{}p(\theta{}|{\rm NSC},\,{}\lambda{}),
\end{equation}
where $\xi{}_1$, $\xi{}_2$, $\xi{}_3$ and $\xi{}_4$ are the mixing fractions of the field (Field), young star cluster (YSC), globular cluster (GC) and nuclear star cluster (NSC) scenarios, defined so that $\xi{}_1 + \xi{}_2 + \xi{}_3 + \xi{}_4 = 1$. Based on this definition, the mixing fraction for each channel is the fraction of  merger events associated with that specific channel.


Figure~\ref{fig:mix_frac} shows the posterior probability distribution of the mixing fractions for the four different channels and for a selection of our models. The results show large variations from one model to another.  The strong fluctuations of $\xi_i$ from one model to another indicate that the mixing fractions are extremely sensitive to the hyper-parameters $\lambda$. Considering the large uncertainties on the astrophysical models, different assumptions about the 1g BBH mass function or other key  parameters  deeply affect the values of $\xi{}_i$. However,  
 there is one common feature: the results of GWTC-2 support the co-existence of multiple channels. In fact, the median value of the mixing fraction is significantly higher than zero for  at least two of the four channels in each specific model. For example, the posterior distributions of the mixing fraction of the NSC, YSC and Field channels peak at values significantly larger than zero in the A5F05, LONG\_DELAY, SHORT\_DELAY, HIGH\_SPIN, BROAD\_VESC,  and NARROW\_VESC models. This result is in agreement with previous work \citep{wong2021,zevin2021,callister2021,bouffanais2021}.

Overall, the Field model seems to be associated with the higher values of the mixing fraction, with median values between $\xi{}_1\sim{0.3}$ (HIGH\_SPIN model) and $\xi{}_1\sim{0.85}$ (HIGH\_MASS model). The rates  and the mass function are  two key ingredients here: the predicted rates are higher for the Field than for the other channels in all our models and the mass function of  Field BBHs has a preference for low values; the BBH population inferred from GWTC-2, after correcting for detection biases, is better represented by a mixture model in which Field binaries give a  substantial contribution. The HIGH\_SPIN case is the one that maximally "penalizes" the Field case, because of an excess of large positive values of $\chi_{\rm eff}$ in this channel. This is the main reason why YSCs and NSCs are expected to contribute significantly to the overall population in this specific model.

In most models, GCs are associated with low mixing fractions. 
This mostly happens because NSCs "work" better than GCs to explain the most massive events (like GW190521) and thus are preferred by our formalism. In the SMALL\_M2 case, both GCs and NSCs are associated with low mixing fractions because they tend to predict too many massive primary BHs with very low mass ratios.

The details of these results might strongly depend on the star formation rate and metallicity evolution model, which can deeply change the merger rate \citep{neijssel2019,tang2019,santoliquido2021,broekgaarden2021}. We will investigate the impact of these quantities in a follow-up study. 
 
\section{Discussion of the main caveats}

We presented a new model that can be used to rapidly simulate hierarchical mergers in different environments (NSCs, GCs and YSCs), exploring a broad parameter space (e.g. progenitor's metallicity, binary evolution parameters such as $\alpha$ and $f_{\rm MT}$, escape velocity from the parent star cluster, delay times and 1g spin distribution). The treatment of dynamical pairing of $N$g BBHs is still approximate: we assume that the retained merger remnants find a new companion and merge over a timescale $dN/dt\propto{}t^{-1}$. This is in agreement with the findings of \cite{antonini2019}, but could be improved with an analytic treatment of dynamical hardening \citep{mapelli2021}. Furthermore, we assume that BHs can only be ejected by relativistic kicks, i.e.  neglect dynamical recoil via close encounters. Finally, we assume that the star cluster does not evolve with time: it has a constant escape velocity. As shown in previous work \citep{breen2013a,breen2013b,moerscher2015,wang2020}, the properties of the star cluster might significantly change  with time and the growth of an IMBH is strongly linked to the evolution of the host star cluster. For example, if we assume constant cluster mass, the half-mass ratio is expected to grow as $r_{\rm h}\propto{}t^{2/3}$ and the escape velocity to decrease with time as $v_{\rm esc}\propto{}t^{-1/3}$ \citep{henon1965}. These two effects might slow down or even suppress the growth of an IMBH in the late evolutionary stages  \cite{antonini2019}. 

In our fiducial model, we assume that the stellar binaries which give birth to 1g BHs are primordial binaries and are not ionized by dynamical interactions. This assumption is motivated by the properties of such binaries. A BBH merger progenitor has an initial binding energy
\begin{equation}
E_b\sim{}6\times{}10^{49}\,{}{\rm erg}\,{}{\rm s}^{-1}\,{}\left(\frac{m_1}{50\,{}{\rm M}_\odot}\right)\,{}\left(\frac{m_2}{50\,{}{\rm M}_\odot}\right)\,{}\left(\frac{1000\,{}{\rm R}_\odot}{\mathcal{A}}\right),
\end{equation}
where  $\mathcal{A}$ is the initial semi-major axis. The typical kinetic energy of a star in a star cluster is
\begin{equation}
E_K\sim{}10^{47}\,{}{\rm erg}\,{}{\rm s}^{-1}\,{}\left(\frac{\langle{m}\rangle}{1\,{}{\rm M}_\odot}\right)\,{}\left(\frac{\sigma{}_{\rm SC}}{100\,{}{\rm km}\,{}{\rm s}^{-1}}\right)^2,
\end{equation}
where $\langle{m}\rangle$ is the average stellar mass in the cluster and $\sigma{}_{\rm SC}$ is the velocity dispersion. 
In the example, we consider an extremely high velocity dispersion $\sigma{}_{\rm SC}=100$ km s$^{-1}$. Hence, binaries that will produce BBH mergers are hard binaries even in the most extreme star clusters and should survive ionization. This assumption breaks in the immediate vicinity of a super-massive BH. 
For example, inside the influence radius of a supermassive BH with mass $m_{\rm BH}=10^6$ M$_\odot$, the typical velocities are $\sim{}120\,{}{\rm km}\,{}{\rm s}^{-1}\,{}(a/0.01\,{}{\rm pc})^{-1/2}\,{}(m_{\rm BH}/10^6\,{}{\rm M}_\odot)$. In this extreme case, even BBHs and their stellar progenitors might be soft binaries and might be broken. On the other hand, dynamical hardening might also be very effective as the BBH gets closer to a supermassive BH by dynamical friction, allowing the BBH to avoid ionization and even speeding up its merger \citep{arcasedda2020b}.  

Here, we make no assumptions about the formation of NSCs. If some of them, if not all, are formed by the hierarchical assembly of GCs \citep{tremaine1975,capuzzo1993}, this might have a crucial impact on the population of BBHs. In fact, the GCs might already be  depleted of merger remnants (because of their relatively low escape velocity) before merging to build up a NSC. Moreover, we neglect the AGN disk formation channel \citep{mckernan2012,stone2017,bartos2017,mckernan2018,yang2020}. Including the physics of AGN disks can boost the contribution of galactic nuclei to the total merger rate and to $N$g mergers. AGN disk physics can further speed up the pairing and merger of our BHs.  We will include the AGN disk scenario in future work.

Arca Sedda et al. (2020, \cite{arcasedda2020}) found remnant masses only up to $\sim{}200$ M$_\odot$, significantly lower than the results presented here for most models. The main reason for this difference is that \cite{arcasedda2020} fixed the escape velocity from NSCs to $v_{\rm esc}=100$ km s$^{-1}$ and did not change this parameter. Our results are consistent with other models (e.g. \cite{antonini2019}), where higher values of $v_{\rm esc}$ are explored. This result is remarkable when considering that \cite{antonini2019} adopt a more accurate model for dynamical interactions than the one presented here. Hence, escape velocities are the key ingredient to understand the mass spectrum of BHs in NSCs.

Finally, we include a simple redshift dependence based on the $f_i(t)$ functions. Alternative redshift dependencies can be obtained by changing $f_i(t)$. For example, if we assume $f_{\rm NSC}=0.1$ (constant with redshift), we obtain an upper limit to the  merger rate density associated with NSCs, because they are unlikely to contribute to 10\% of the overall cosmic star formation rate. Under such extreme assumption, the local BBH merger rate density from NSCs is $\mathcal{R}_{\rm NSC}\approx{}7-10$ Gpc$^{-3}$ yr$^{-1}$, i.e. approximately a factor of 10 higher than the models we presented in Figures~\ref{fig:MRD} and \ref{fig:MRD_other}.

\section{Summary}

Hierarchical mergers in dynamical environments can lead to the formation of BHs with mass higher than the limits imposed by pair instability, core-collapse supernovae and stellar evolution theory. Here, we  presented a fast semi-analytic method to draw the main properties (masses, spins, merger rate) of hierarchical BBHs, while probing the relevant parameter space.

In our models, NSCs are the dominant environment for the formation of hierarchical BBHs. In our fiducial model (A5F05), primary BHs with mass up to $\sim{}10^3$ M$_\odot$ can form in NSCs, while the maximum primary BH mass is $\sim{}100$ M$_\odot$ for both  GCs and YSCs.

We find that the mass distribution of 1g BBHs has a crucial impact on the mass distribution of $N$g BHs with $N>1$. The metallicity of the progenitor is a key ingredient to shape the distribution of $N$g BBHs, because it affects both the number and the maximum mass of BBHs. The common envelope $\alpha{}$ parameter and the accretion efficiency $f_{\rm MT}$ also play a role, with smaller values of $\alpha{}$ leading to higher merger rates and higher values of $f_{\rm MT}$ leading to more top-heavy BH mass functions.

If BHs with mass in the pair instability gap are allowed to form by stellar mergers \citep{dicarlo2020}, the mass distribution of $N$g BBHs is skewed toward significantly larger masses  (HIGH\_MASS model).  
Primary BH masses up to a few $\times{}10^4$ M$_\odot$ can be obtained in NSCs if only $N$g$-$1g mergers are allowed to take place, i.e. if we prevent the secondary BH from being a merger remnant itself (SMALL\_M2 model). The main reason is that relativistic kicks are smaller if the mass ration $q=m_2/m_1$ tends to zero.

The escape velocity of the parent star cluster ($v_{\rm esc}$) is probably the most important parameter to set the maximum BH mass. If we assume that the distribution of escape velocities from NSCs is $\log_{10}(v_{\rm esc}/{\rm km}\,{}{\rm s}^{-1})=2\pm{}0.3$ (BROAD\_VESC model), BHs with mass up to $\sim{}10^6$ M$_\odot$ are allowed to form in the NSCs with the highest escape velocities. This result is consistent with \cite{antonini2019} and \cite{fragione2020}.

While BBHs in GCs and YSCs do not exceed the 5th and the 3rd generation, respectively,  we expect at least 10 different BBH generations in NSCs. This number grows up to a few thousands if $N$g$-$1g BBHs are the only way to produce hierarchical mergers (SMALL\_M2 model).

In our fiducial model, the fraction of $N$g BBHs is $f_{>1g}\sim{}0.15$ in NSCs, which lowers to $6\times{}10^{-3}$ in GCs and $\sim{}10^{-4}$ in YSCs. In the most optimistic case (i.e. when low spins are assumed for 1g BHs), $f_{>1g}\sim{}0.5$, 0.1, 0.01 for NSCs, GCs and YSCs, respectively. In the most pessimistic case (i.e. when high spins are assumed), $f_{>1g}\sim{}0.08$, $2\times{}10^{-3}$ and  $4\times{}10^{-5}$ for NSCs, GCs and YSCs, respectively.

BHs in the pair instability mass gap and IMBHs can form via hierarchical mergers. Their fraction is strongly suppressed at high metallicity. At  $Z=0.0002$ and in our fiducial model, the fraction of BBH mergers with primary BH mass in the pair instability gap is $f_{\rm PISN}\sim{}7\times{}10^{-3}$, $3\times{}10^{-4}$ and $5\times{}10^{-6}$ 
in NSCs, GCs and YSCs, respectively. In our fiducial model, the fraction of BBH mergers with primary BH mass in the IMBH regime is $f_{\rm IMBH}\sim{}5\times{}10^{-4}$ in NSCs, while we do not find any IMBH mergers in either GCs or YSCs. These fractions are significantly higher in the SMALL\_M2 and  HIGH\_MASS models (Figure~\ref{fig:ngen_PISN}).

The local BBH merger rates in our models range from $\sim{}10$ to $\sim{}60$ Gpc$^{-3}$ yr$^{-1}$, but $N$g BBHs in NSCs only account for  $10^{-2}-0.2$ Gpc$^{-3}$ yr$^{-1}$ in our models. If we assume that 10\% of all stars form in NSCs, we find a robust upper limit $\sim{}7-10$ Gpc$^{-3}$ yr$^{-1}$ for the local merger rate density of $N$g BBHs in NSCs.

We compare our models against LIGO--Virgo data from the second gravitational wave transient catalog (GWTC-2, \citep{abbottO3a,abbottO3apopandrate}), by estimating the mixing fractions of the four considered channels. Even if the mixing fractions are wildly affected by model hyper-parameters, our analysis suggests that more than one channel is needed to explain the observed population from GWTC-2. 
This result confirms that the BBHs observed by the  LIGO--Virgo collaboration  likely are a combination of several different channels and opens new perspectives for the study of BBH formation.



\authorcontributions{Conceptualization, M.M., M.A.S. and M.C.A.; methodology, M.M., F.S. and Y.B.; software, M.M., F.S. and Y.B.; writing---original draft preparation, M.M. and A.B.; project administration, M.M.; funding acquisition, M.M., M.C.A. and M.A.S.. All authors have read and agreed to the published version of the manuscript.}

\funding{M.M., F.S., Y.B., and A.B. acknowledge financial support from the European Research Council for the ERC Consolidator grant DEMOBLACK, under contract no. 770017. M.M. and M.C.A. acknowledges financial support from the Austrian National
Science Foundation through FWF stand-alone grant P31154-N27. 
M.A.S. acknowledges financial support from the Alexander von Humboldt Foundation for the research program ``The evolution of black holes from stellar to galactic scales'', the Volkswagen Foundation Trilateral Partnership for project No. I/97778 ``Dynamical Mechanisms of Accretion in Galactic Nuclei'', and the Deutsche Forschungsgemeinschaft (DFG, German Research Foundation) -- Project-ID 138713538 -- SFB 881 ``The Milky Way System''.}



\dataavailability{The data underlying this article will be shared on reasonable request to the corresponding authors.} 

\acknowledgments{We thank Eugenio Carretta for useful discussion and we thank the internal referee of the LIGO--Virgo collaboration, Fabio Antonini, for his suggestions, which helped us improve this work.}

\conflictsofinterest{The authors declare no conflict of interest. The funders had no role in the design of the study; in the collection, analyses, or interpretation of data; in the writing of the manuscript, or in the decision to publish the~results.} 



\abbreviations{Abbreviations}{
The following abbreviations are used in this manuscript:\\

\noindent 
\begin{tabular}{@{}ll}
  AGN & active galactic nucleus\\
  BBH & binary black hole\\
  BH & black hole\\
  GC & globular cluster\\
  GW & gravitational wave\\ 
  GWTC-2 & second gravitational wave transient catalog\\
  IMBH & intermediate-mass black hole\\
  NSC & nuclear star cluster\\
  YSC & young star cluster
\end{tabular}}




\end{paracol}
\reftitle{References}


\externalbibliography{yes}
\bibliography{bibliography}

\end{document}